\documentclass[]{spie}  

\usepackage{amsmath,amsfonts,amssymb}
\usepackage{graphicx}
\usepackage[colorlinks=true, allcolors=blue]{hyperref}

\usepackage{algorithm}
\usepackage{algpseudocode} 

\newcommand{\Qhat}{\hat{Q}}
\newcommand{\Rhat}{\hat{R}}

\title{Quasi Real-Time Autonomous Satellite Detection and Orbit Estimation}

\author[a]{Jarred Jordan}
\author[a]{Daniel Posada}
\author[a]{Matthew Gillette}
\author[a]{David Zuehlke}
\author[a]{Troy Henderson}
\affil[a]{Space Technologies Laboratory, Embry-Riddle Aeronautical University, 1 Aerospace Blvd., Daytona Beach, FL, 32114, United States}

\authorinfo{Further author information: (Send correspondence to Troy Henderson, hendert5@erau.edu)\\ T.H. E-mail: hendert5@erau.edu}

\begin{document} 
\maketitle

\begin{abstract}
A method of near real-time detection and tracking of resident space objects (RSOs) using a convolutional neural network (CNN) and linear quadratic estimator (LQE) is proposed. Advances in machine learning architecture allow the use of low-power/cost embedded devices to perform complex classification tasks. In order to reduce the costs of tracking systems, a low-cost embedded device will be used to run a CNN detection model for RSOs in unresolved images captured by a gray-scale camera and small telescope. Detection results computed in near real-time are then passed to an LQE to compute tracking updates for the telescope mount, resulting in a fully autonomous method of optical RSO detection and tracking. 
\end{abstract}


\section{Introduction}
As space missions and launches become more frequent and accessible, it is important to optimize Space Domain Awareness (SDA), debris tracking, and resident space object (RSO) detection capabilities. The accurate identification and tracking of satellites in space imagery are crucial for a range of applications, including safety, reconnaissance, contingency planning, situational awareness, and debris removal. With the growing number of satellites in orbit, particularly due to new super constellations like SpaceX's StarLink and OneWeb, there is an urgent need for rapid and efficient methods to process SDA observations. Current estimates suggest that there are around 700,000 objects larger than 1cm in orbit, the vast majority of which are uncontrolled debris. Various techniques have been developed for processing optical observations of RSOs, such as gross motion analysis across frames, inertial frame angular observations, and template matching. Numerous studies have investigated these methods, including those by Schildknecht et al. (2007), Virtanen (2016), Sease et al. (2015), Zuehlke et al. (2020, 2021), and Lovell et al. (2021).\cite{schildknechtOpticalSurveys_2007,Virtanen2016StreakDetection-fz,Brad_Sease2015-ge,Zuehlke2020-em,Zuehlke2021-uv,Virtanen2016-we,Zuehlke2020EndtoEnd-uz,Alan_Lovell2021Processing-dm} 

With the steady advances in object detection models in machine learning and computer performance in general, it has become easier to conduct complex tasks on low-cost embedded devices for space applications. In an effort to improve the efficiency of ground-based observation and tracking of RSOs, it is advantageous to utilize these developments to enhance the detection capability compared to a human-in-the-loop. This paper discusses the application of an algorithm containing a neural network detection model\cite{jordan2022satellite} to rapidly detect and track uncatalogued RSOs in near real-time on a portable system that can be attached to a telescope. Detected RSOs from the CNN are passed to a linear quadratic estimator (LQE) to provide tracking updates to a telescope/camera system to enable repeated observations. Once a number of observations have been compiled, measurements are available to begin tracking the RSO without any prior information. The detection model will be running on a small Raspberry Pi system, in tandem with a small ground-based commercial telescope and tracking mount. 

A limitation of many current SDA tracking algorithms is that they require \textit{a-priori} information available to acquire and track a given RSO. The real-time tracking algorithm presented in this research requires no \textit{a-priori} information about an RSO to begin tracking. Once an RSO has been detected by the CNN, the tracking process begins, and repeated observations are able to be captured. Observations can then be passed to an orbit determination routine to provide an estimate of the observed RSOs position and velocity. 

\section{Background}

\subsection{Machine Learning}
Machine learning (ML) is a field of computer science that focuses on developing algorithms and statistical models that allow computers to learn from data and make predictions or decisions without being explicitly programmed. Convolutional neural networks (CNNs) are a type of deep learning model that has revolutionized the field of computer vision. They are specifically designed to process image data by automatically learning and identifying patterns in the data through convolutional layers, pooling layers, and fully connected layers. CNNs have been widely applied in various image recognition tasks, such as object detection, face recognition, and medical image analysis. Their ability to learn from large datasets, combined with advancements in hardware and software technology, has led to significant improvements in the accuracy of image recognition tasks in recent years.\cite{gu2018recent,li2021survey}

\begin{figure}[!htb]
    \centering
    \begin{minipage}{0.48\textwidth}
        \centering
        \includegraphics[width=0.9\textwidth]{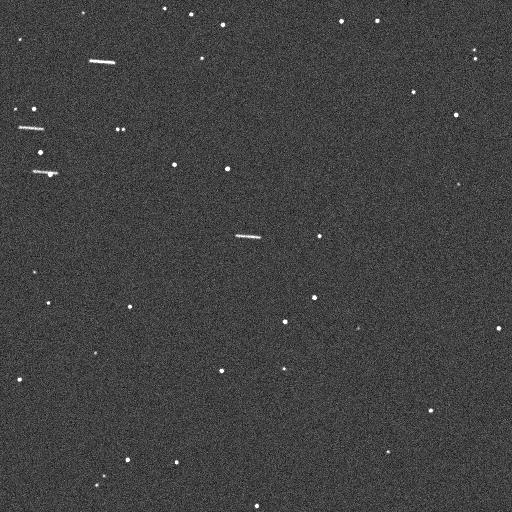}
    \end{minipage}%
    \begin{minipage}{0.48\textwidth}
        \centering
        \includegraphics[width=0.9\textwidth]{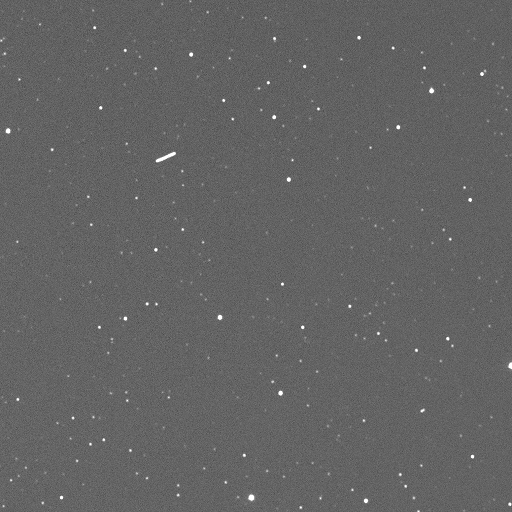}
    \end{minipage}
    \caption{Sample training images provided to the network; synthetic imagery (left); real imagery (right).}
    \label{fig:SatSim}
\end{figure}

In order to improve the accuracy of RSO detection by the CNN, multiple images with RSOs are labeled under various tracking conditions. As illustrated in Figure \ref{fig:SatSim}, stars and RSOs can appear either as defined ellipses (point-sources) or streaks, making it necessary to ensure that the model architecture is robust to different image capture settings. For this particular purpose, the network was trained to detect the RSOs as streaks. The CNN architecture is able to find specific features and gradients in the image that are unique to the shape and brightness of the RSO while ignoring the noise and stars in the image.

The proposed architecture is based on the state-of-the-art deep learning model MobileNetV2.\cite{Sandler2019MobileNetV2IR} This network comprises multiple layers that utilize convolutions and bottlenecks to map the implicit information in the image into various higher dimensions. This mapping enables the model to identify the unique patterns that differentiate an RSO from a star \cite{karadal2021automated}. A secondary reason but significant for this architecture is the number of hyperparameters and network size as it is originally intended to run on mobile phones. A key feature of this particular model implementation is the inclusion of a single shot detector (SSD) framework, which provides several advantages over other pre-trained object detection frameworks.\cite{LiuWei2016SSSM, posada2022nn} Furthermore, SSD-reinforced models are currently the only models that are compatible with TensorFlow Lite (TFLite) conversion requirements. This is crucial, as converted models are optimized for mobile inference on embedded systems, resulting in additional improvements to resource overhead and inference latency when compared to conventional TensorFlow models.

\begin{figure}[!htb]
    \centering
    \begin{minipage}{\textwidth}
        \centering
        \includegraphics[width=\textwidth]{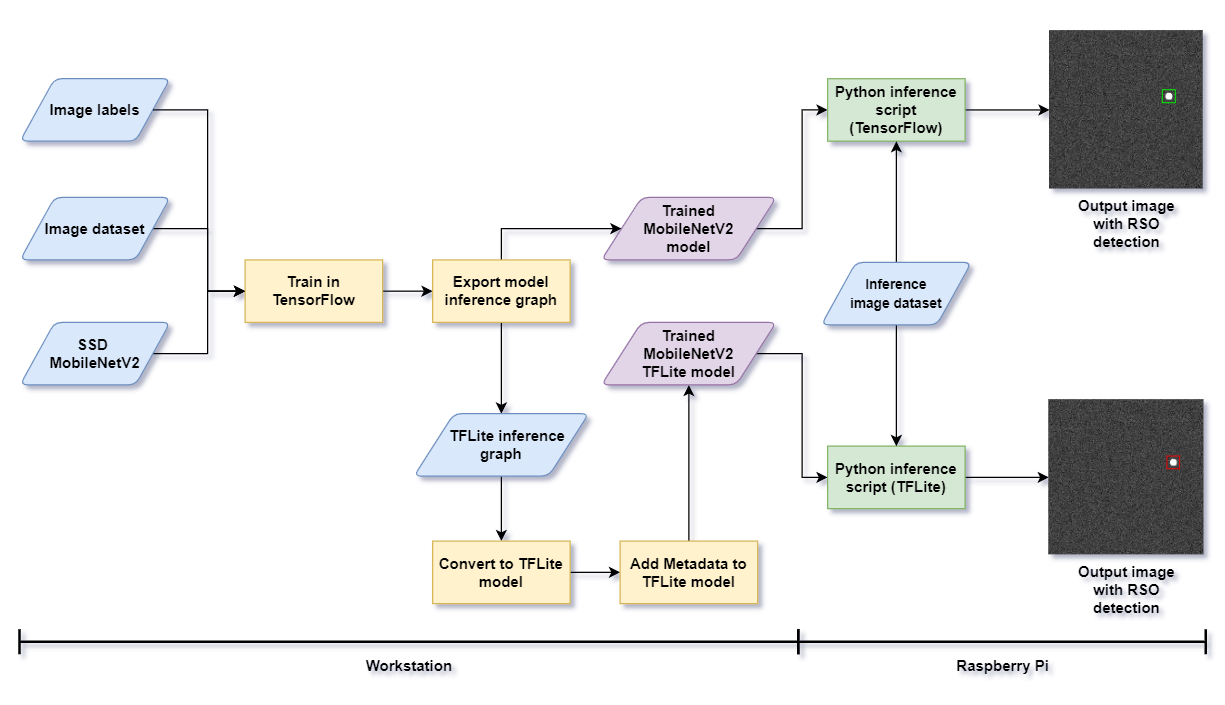}
    \end{minipage}%
    \caption{General procedure for training and image inference for RSO detection using MobileNetV2.\cite{jordan2022satellite}}
    \label{fig:Flowchart:nn}
\end{figure}

The workstation used for training the detection models contained two Intel Xeon Silver 4214 processors, 128 GB of RAM, and an NVIDIA Quadro RTX 4000. For the mobile application of the model, the tests utilized a Raspberry Pi 4 with 4 GB of system RAM and a slightly overclocked CPU which operates at 1800 MHz. Additionally, the Raspberry Pi was operated at standard ambient conditions and did not utilize any external computation modules or extra cooling. For more information about the database and training as seen in Figure \ref{fig:Flowchart:nn}, the reader is referred to previous work by the authors which explains in detail the training process and network details.\cite{jordan2022satellite}

\subsection{Estimation Systems}
Estimators for tracking centroids on an image have existed since the middle of the 20th century. One of the earliest methods for determining the state of a dynamic system that was extensively utilized was the Kalman filter, which was created by Rudolf Kalman in 1960.\cite{welch1995introduction} It is a recursive method that estimates the system's status and forecasts future values using a sequence of measurements taken over time. The linear quadratic estimator (LQE), a replacement for the Kalman filter, was introduced in the 1970s.\cite{morris1976kalman} The LQE is a more versatile framework that can handle systems with non-linear dynamics and supports the use of various cost functions.\cite{anderson2007optimal} These estimators have undergone several modifications and enhancements over time, including the extended Kalman filter (EKF) and the unscented Kalman filter (UKF). In a number of applications, including image and object tracking, navigation, and control systems, these estimators have evolved into crucial tools.

A target's location in an image can be estimated using the mathematical framework known as LQE. The LQE can be used in the context of image processing to determine the centroid of an object in a collection of images taken over time. The LQE calculates the target's position based on a set of observations or the values of the pixels in the image. The LQE uses a cost function to determine the best estimate of the target's position while assuming that the observations are tainted by noise. The cost function can be chosen to maximize estimate likelihood or minimize mean squared error depending on the particular problem. Since the LQE can deal with non-linear dynamics and non-Gaussian noise, it has been widely used in computer vision applications like object tracking and image registration. An example of an LQE implemented to identify the centroid and track the movement of a robotic manipulator is described by Daniilids (1998).\cite{daniilidis1998real} Other common methods for tracking objects make use of optical flow to detect the motion.\cite{zuehlke2022autonomous} The LQE framework will be adapted from tracking object centroids in an image to tracking inertial angular positions of an RSO as it transits the night sky. 

\section{Methodology}
The details of the real-time tracking algorithm are discussed in this section. First, the overall algorithm will be presented, followed by details on the LQE estimator, embedded-system architecture, and image processing. 
\subsection{Real-Time Detection Algorithm}
\begin{figure}[!htb]
    \centering
    \includegraphics[width=\textwidth]{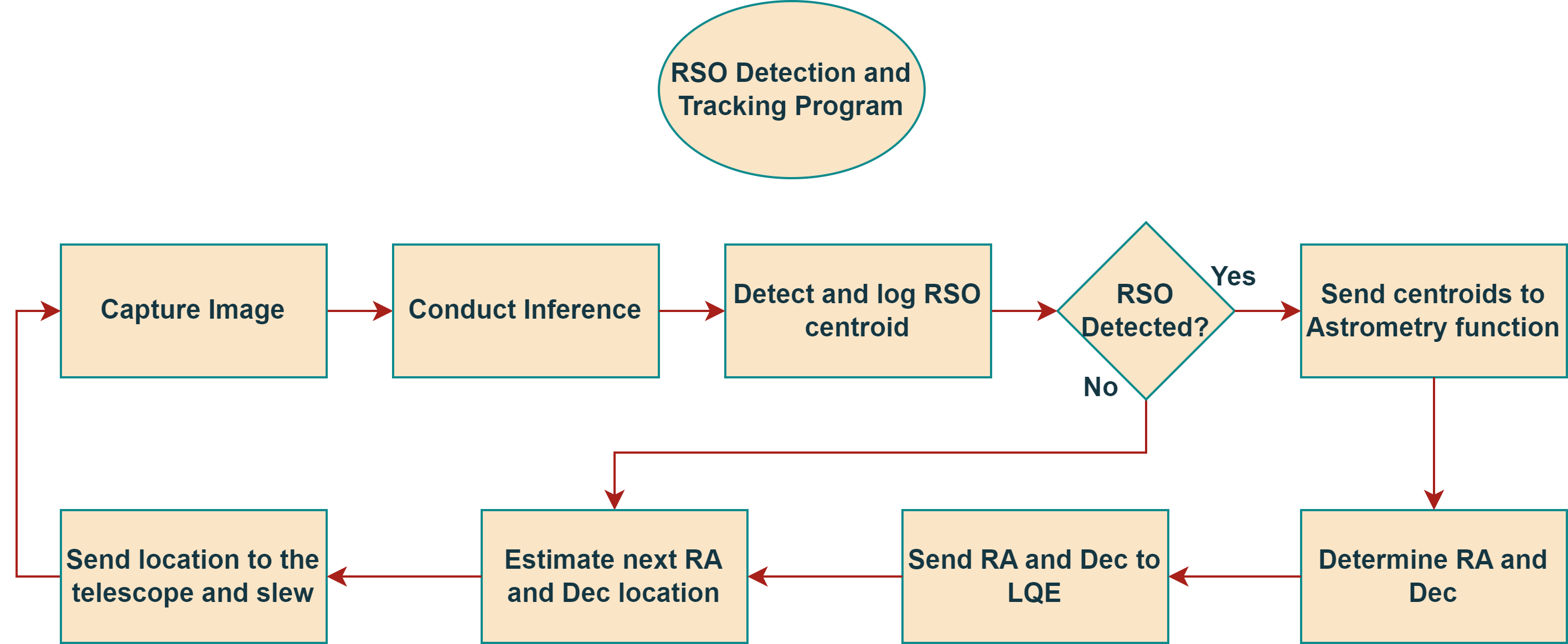}
    \caption{High-level program architecture.}
    \label{fig:STARLITE}
\end{figure}

This research proposes an autonomous algorithm for tracking RSOs without any prior information of their orbit. The algorithm is completely autonomous and requires no human-in-the-loop interaction to acquire and track an RSO. The heart of the detection routine is the CNN used to detect an RSO in an image. Detected RSO locations are then used as measurements for an LQE tracker to predict the RSOs future location. The overall RSO tracking algorithm flow is pictured in Fig. \ref{fig:STARLITE}. 

The program begins by capturing images of the sky tracking at a sidereal rate (the rate of the earth's rotation) such that stars remain fixed in the image and any RSOs appear as streaks in the image. After each image is captured, the inference is run using the trained CNN to detect if an RSO is present. If an RSO is not detected, another image is taken, a process that is repeated until a positive detection is made. Once an RSO is detected by the CNN, the image is then plate-solved in by \texttt{astrometry} (discussed in Section \ref{sec:astrometry}) to provide the inertial Right Ascension (RA) and Declination (DEC) coordinates of the detected RSO. A second image is then taken and inference is run to detect the RSO and provide a second measurement in order to initialize the LQE to predict the RSO's future location in the sky. With two measurements, the LQE is initialized and the RSOs location in the sky is predicted by propagating the estimated state forward to the next image capture time. The telescope is then sent a slew command to move to the estimated position in the sky. 

Once the telescope has finished slewing, another image is captured and inference is run to detect the RSO at its new location in the sky. If the RSO is detected, the image is plate-solved with \texttt{astrometry} as before to provide an RA, DEC measurement. The measurement is then used to update the LQE filter. Once the filter has been updated with the newly detected RSO position, the RSO position is again propagated forward in time and the telescope is slewed to capture another image to continue with closed-loop tracking. During each iteration, if the RSO was successfully detected by the CNN, the LQE is updated with the measurement. If for some reason the CNN failed to detect the RSO, no measurement update is provided for that iteration, and the RSO position is predicted using the LQE filter's propagate feature. 

Further details into the tracking algorithm are given in Alg. \ref{alg:real:time:detection}. Details for each of the operations are explained in the following sections. Once fully written, the program was implemented in a complete in-the-loop demonstration with a telescope and a camera connected to a Raspberry Pi. 
\clearpage
\begin{algorithm}[hbt!]
\caption{Quasi-real time RSO detection and tracking algorithm.}\label{alg:real:time:detection}
\begin{algorithmic}[1] 
\Require Initialize TF network, initialize \texttt{astrometry}
\For{$k = 1$ to $M$}
    \State Acquire image: $I_k$
    \State Process image: $I_p \gets $\texttt{process\_image}$(I_k)$
    \State Run inference on processed image $I_p$
    \If{Confidence is $>$ $\tau$}
       \State RSO Centroid $\gets$ center of bounding box of CNN detection
       \State Plate solve image: $WCS \gets$ \texttt{plate\_solve}$(I_p)$
       \State Compute RA, DEC measurement from image registration solution $WCS$: 
       \State $\mathbf{y}_k \gets (\alpha_k,\delta_k) \gets$ \texttt{convert\_to\_world\_coords}$(x,y)$
    \Else
        \State No RSO detected return to the top and take another image
    \EndIf
    \If{$EPOCH == 1$}
        \State Initialize LQE with two RSO measurements
    \Else 
        \State Update the filter with the current measurement
    \EndIf
    \State Propagate RSO location using LQE
    \State Slew telescope to the predicted location
    \State Go back to the top and take another image
\EndFor
\State \Return Tracked locations of satellites
\end{algorithmic}
\end{algorithm}

\subsection{LQE Estimator} 
An overview of the LQE system used to estimate the location of RSOs for tracking is presented in this section. Note that the LQE described in this section will be operating in discrete time. The beginning point of an LQE is to model the system and observation dynamics as a linear system as shown in Eqs. \ref{eqn:xhatdot} and \ref{eqn:yk}. The estimated states are given by the vector $\hat{\mathbf{x}}$ and the estimated measurement vector is given by $\mathbf{y}_k$. The standard linear system format is followed where the $A$ matrix is the system state matrix, $B$ is the control input matrix, $C$ is the observation matrix, and $D$ is the control observation matrix. Note that for the system considered here, the control input is not modeled and both the $B$ and $D$ matrices are set to zero matrices of appropriate sizes. 
\begin{equation}
    \dot{\hat{\mathbf{x}}} = A \hat{\mathbf{x}} + B \mathbf{u} \label{eqn:xhatdot}
\end{equation}
\begin{equation}
    \mathbf{y}_k = C \hat{\mathbf{x}} + D \mathbf{u} \label{eqn:yk}
\end{equation}
The dynamics for the system are the dynamics of an RSO moving in inertial (RA, DEC) space, which have been shown to be linear over short time spans.\cite{Zuehlke2020EndtoEnd-uz,Joseph_D_Anderson-ej} The system estimated states at time $t_k$ are given by the vector $\hat{\mathbf{x}}$ and are defined to be the right ascension, right ascension rate, declination, and declination rate $\hat{\mathbf{x}} = \begin{bmatrix} \alpha & \dot{\alpha} & \delta & \dot{\delta}    \end{bmatrix}^T$ of an observed RSO. 

Assuming linear dynamics for the observed RSO, the $A$ matrix will be set as shown in Eq. \ref{eqn:Amatrix} to correspond to the state vector shown in Eq. \ref{eqn:xhat}. The measurements for the LQE will be provided from detected RSO positions found by the CNN converted to RA, DEC $(\alpha, \delta)$ angles. Since the observed states are the angles $(\alpha, \delta)$, the $C$ matrix is defined as in Eq. \ref{eqn:Cmatrix} to form the correct measurement vector when multiplied with the estimated state vector.

\begin{equation}
    A = \begin{bmatrix}
        0 & 1 & 0 & 0 \\
        0 & 0 & 0 & 0 \\
        0 & 0 & 0 & 1 \\
        0 & 0 & 0 & 0 
    \end{bmatrix} \label{eqn:Amatrix}
\end{equation}
\begin{equation}
        \hat{\mathbf{x}} = \begin{bmatrix}
        \alpha & \dot{\alpha} & \delta & \dot{\delta}
    \end{bmatrix}^T \label{eqn:xhat}
\end{equation}
\begin{equation}
        C = \begin{bmatrix}
        1 & 0 & 0 & 0 \\
        0 & 0 & 1 & 0 
    \end{bmatrix} \label{eqn:Cmatrix} 
\end{equation}

The LQE filter works through a combination of propagation of the estimated states and a measurement update process. Process and measurement noise are factored into the dynamics of the system through the matrices $\Qhat$ and $\Rhat$ defined below. The process and measurement noise are set based on the expected noise in the system. Where the noise value is set to be $\sigma = 4$ arcseconds (1 arcsecond $= 1/3600$ of a degree). 

\begin{equation}
    \Qhat = \begin{bmatrix}
        500 \sigma & 0 & 0 & 0 \\
        0 & \dfrac{\sigma}{100} & 0 & 0 \\
        0 & 0 & \dfrac{\sigma}{5} & 0 \\
        0 & 0 & 0 & \dfrac{\sigma}{100} 
    \end{bmatrix} \label{eqn:Qhat} 
\end{equation}
\begin{equation}
    \Rhat = \begin{bmatrix}
        2 \sigma^2 & 0 \\
        0 & \sigma^2 
    \end{bmatrix}
\end{equation}

In order to initialize the filter, a set of two angular measurements for an RSO are needed to provide both an angular position and angular velocity first estimate. Once two detections have been made, these first two readings are used to acquire the initial conditions. The initial angles are defined as the angles from the second reading. The initial angular velocities are solved by interpolating 2 lines between the 2 readings, a line for RA and a line for DEC. The slopes for these 2 lines are equal to the velocity between these 2 points. This is used for the initial velocities.

In addition to the filter states, the filter covariance must also be initialized. The filter covariance $P_k^{-}$ is initialized as the $4 \times 4$ identity matrix. The filter estimated states $\hat{\mathbf{x}}_k^{-}$ are initialized as discussed above. Once initialized, the states and covariance are propagated using Eqs. \ref{eqn:xhatminus:prop} and \ref{eqn:Pkminus:prop}. Where the state transition matrix $F$ is given as the matrix exponential of the system matrix $A$ multiplied by the propagation time $\Delta t$. Note that the filter propagate and update steps are indicated by a $-$ and $+$ superscript respectively. For example, the propagated state at time $t_k$ before updating the filter state with a measurement is given by $\hat{\mathbf{x}}_k^{-}$ and once the filter update is applied the state is denoted $\hat{\mathbf{x}}_k^{+}$. The estimated states are propagated to the next measurement time using Eq. \ref{eqn:xhatminus:prop}. The filter covariance is propagated to the next measurement time using Eq. \ref{eqn:Pkminus:prop}. Note that the process noise factors into the covariance propagation. 

\begin{equation}
    \hat{\mathbf{x}}_k^{-} = F \hat{\mathbf{x}}_{k-1}^{+}  \label{eqn:xhatminus:prop}
\end{equation}

\begin{equation}
    P_k^{-} = F P_{k-1}^{+} F^T + Q_{k - 1} \label{eqn:Pkminus:prop}
\end{equation}

\begin{equation}
    F = e^{A \Delta t} \label{eqn:expm}
\end{equation}

\begin{figure}[!htb]
    \centering
    \includegraphics[width=0.65\textwidth]{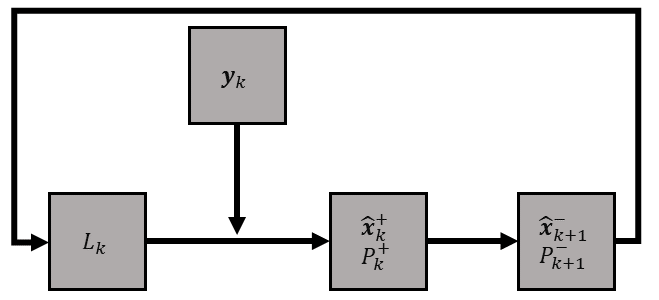}
    \caption{Linear quadratic estimator flow diagram.}
    \label{fig:LQE}
\end{figure}

After propagation and a new measurement from the CNN given by $\Tilde{\mathbf{y}}_k = \begin{bmatrix}
    \alpha_k & \delta_k 
\end{bmatrix}^T$ is available, the filter estimated states and covariance are updated by finding the Kalman gain. The Kalman gain is denoted $L_k$ and is calculated using Eq. \ref{eqn:kalman:gain}. Note that the measurement noise matrix is factored into the Kalman gain calculation. Once the measurement has been read and the gain calculated the estimated states are updated using Eq. \ref{eqn:xhatkplus} and the covariance is updated by Eq. \ref{eqn:Pkplus}. After the update step, the filter states are propagated forward in time to predict the RSO location, and the estimated RA, DEC positions are sent to the telescope. After the next image is captured and another measurement is available the filter states are updated again and the process continues in a loop. A visual representation of the filter loop is shown in Fig. \ref{fig:LQE}. 

\begin{equation}
    L_k =  P_k^{-} C^T \left(C P_k^{-} C^T + R_k \right)^{-1} \label{eqn:kalman:gain}
\end{equation}

\begin{equation}
    \hat{\mathbf{x}}_k^{+} = \hat{\mathbf{x}}_k^{-} + L_k \left(\Tilde{\mathbf{y}} - C \hat{\mathbf{x}}_k^{-}\right) \label{eqn:xhatkplus}
\end{equation}

\begin{equation}
    P_k^{+} = \left(I_{4 \times 4} - L_k C\right) P_k^{-} \label{eqn:Pkplus}
\end{equation}

\subsection{Embedded System Architecture}
\begin{figure}[!htb]
    \centering
    \includegraphics[width=0.9\textwidth]{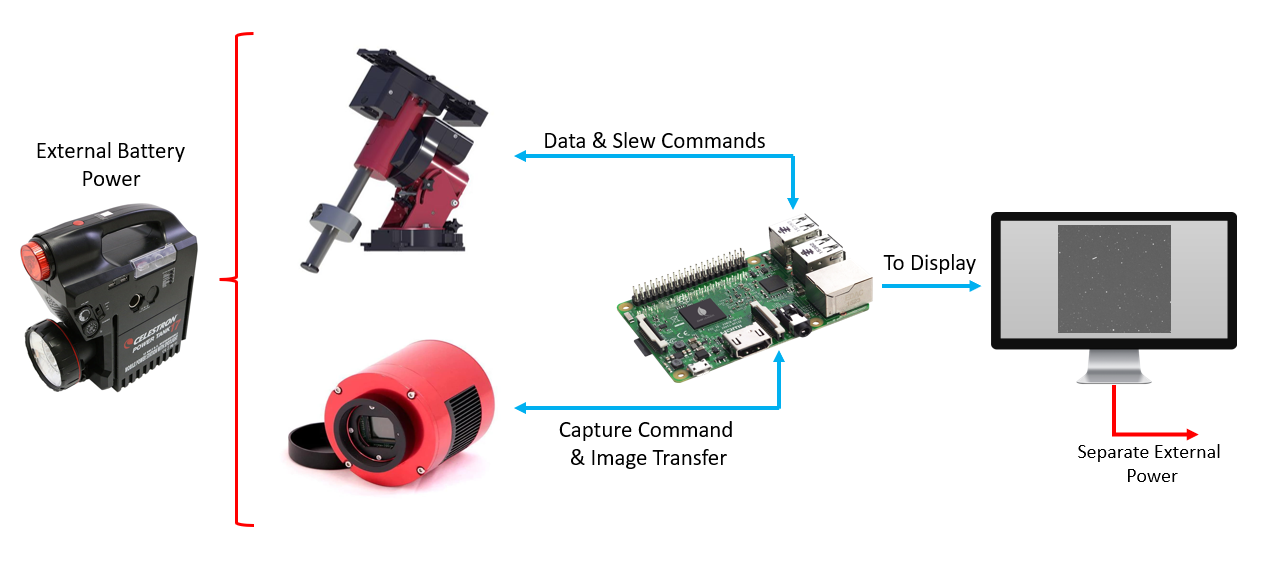}
    \caption{Mobile setup and hardware interaction (telescope tube not shown).}
    \label{fig:setup:diagram}
\end{figure}

\begin{figure}[!htb]
    \centering
    \includegraphics[width=\textwidth]{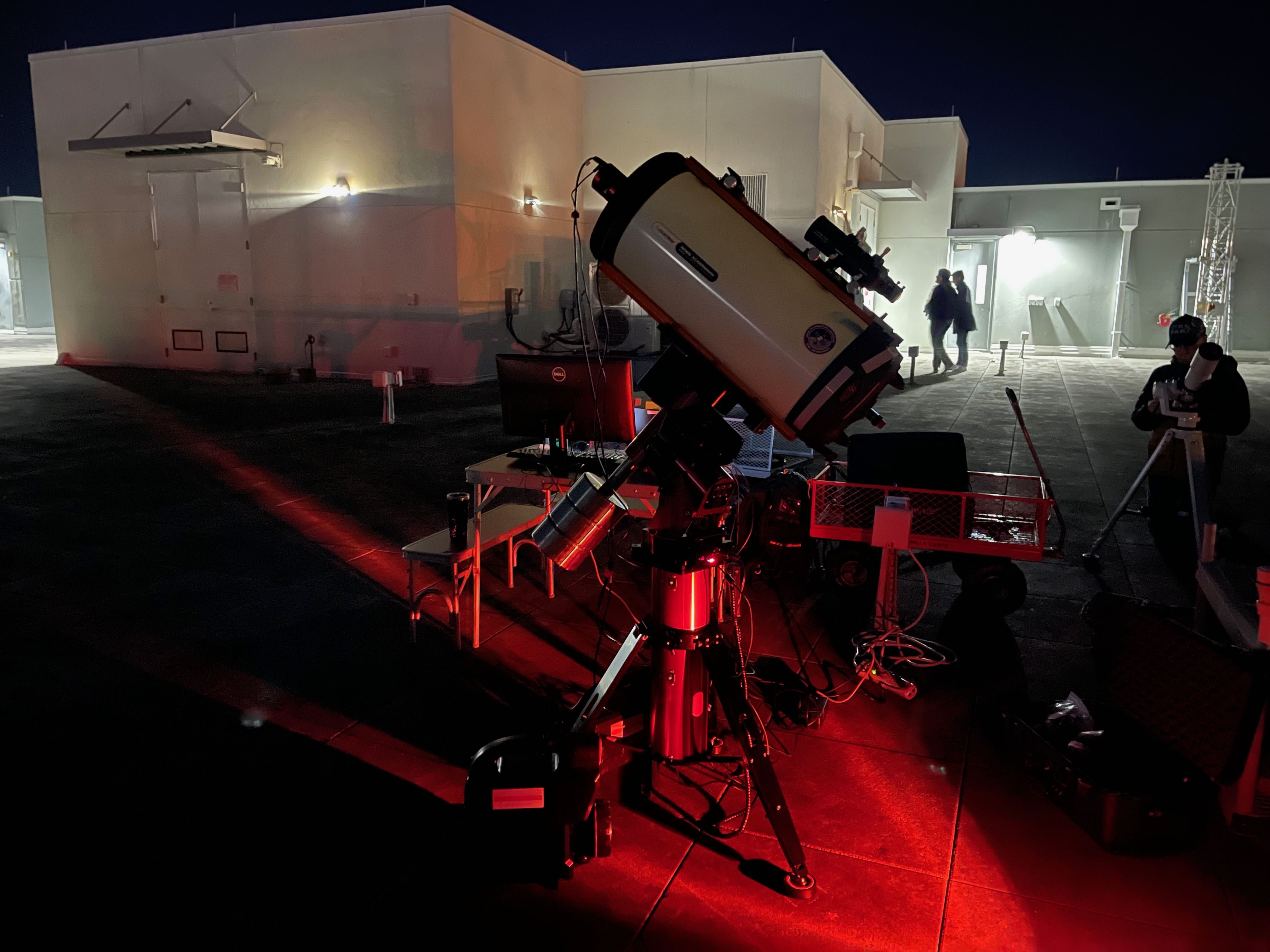}
    \caption{Optical tracking setup at the roof of the MicaPlex building at ERAU.}
    \label{fig:setup:picture}
\end{figure}

In order to achieve the goal of real-time autonomous RSO tracking, the following components were used for testing the algorithm. Figure \ref{fig:setup:diagram} illustrates the setup and the different hardware components used to test the tracking algorithm. An ASI 1600MM-Pro monochrome camera is used for image acquisition. A Software Bisque Paramount MYT mount is used as the base for an $11$in Celestron Rowe Ackerman Schmidt Astrograph (RASA) telescope.\cite{RASA2016-rt} Using a custom INDI\footnote{INDI Library, \url{https://indilib.org/}} driver communication is achieved between the Raspberry Pi and the mount and camera. This allows autonomous control over image acquisition of the images and the commanding of the telescope by sending desired $RA$ and $DEC$ commands. Figure \ref{fig:setup:picture} shows the optical tracking setup utilized for this research on the roof of the MicaPlex building at Embry-Riddle Aeronautical University (ERAU). 

\subsection{Astrometry and Image Localization} \label{sec:astrometry}
In order for an image to provide a useful measurement for tracking a satellite, the angular position of the satellite with reference to the background sky must be found. Finding the orientation of an image with respect to known background stars is accomplished through a process called plate-solving. By plate-solving known star positions are compared to star positions in an image in unique patterns often referred to as ``asterisms.'' The detected star patterns are then compared to catalog positions to recover the image orientation and the center of the field of view. A Python implementation of the \textit{astrometry.net} plate-solving software was used to perform image registration.\cite{Marcireau2022-ev} 

The \texttt{astrometry} package works by converting a set of $x,y$ coordinate pairs representing stars in an image into unique patterns to compare against the built-in star catalogs. Star centroids are computed using intensity-weighted centroids.
\subsection{Image Processing}

\begin{figure}[!htb]
    \centering
    \begin{minipage}{0.5\textwidth}
        \centering
        \includegraphics[width=0.9\textwidth]{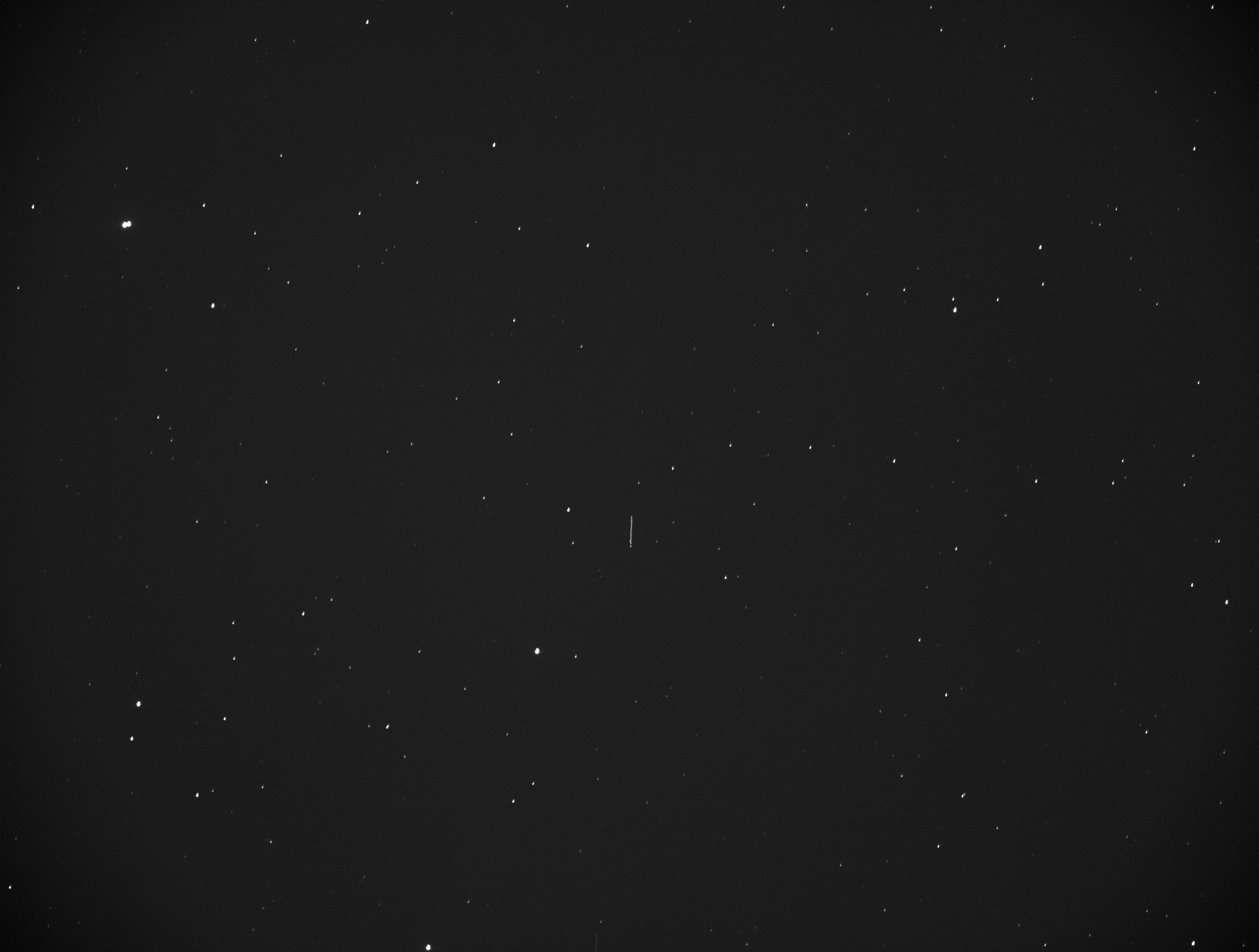}
    \end{minipage}%
    \begin{minipage}{0.5\textwidth}
        \centering
        \includegraphics[width=0.9\textwidth]{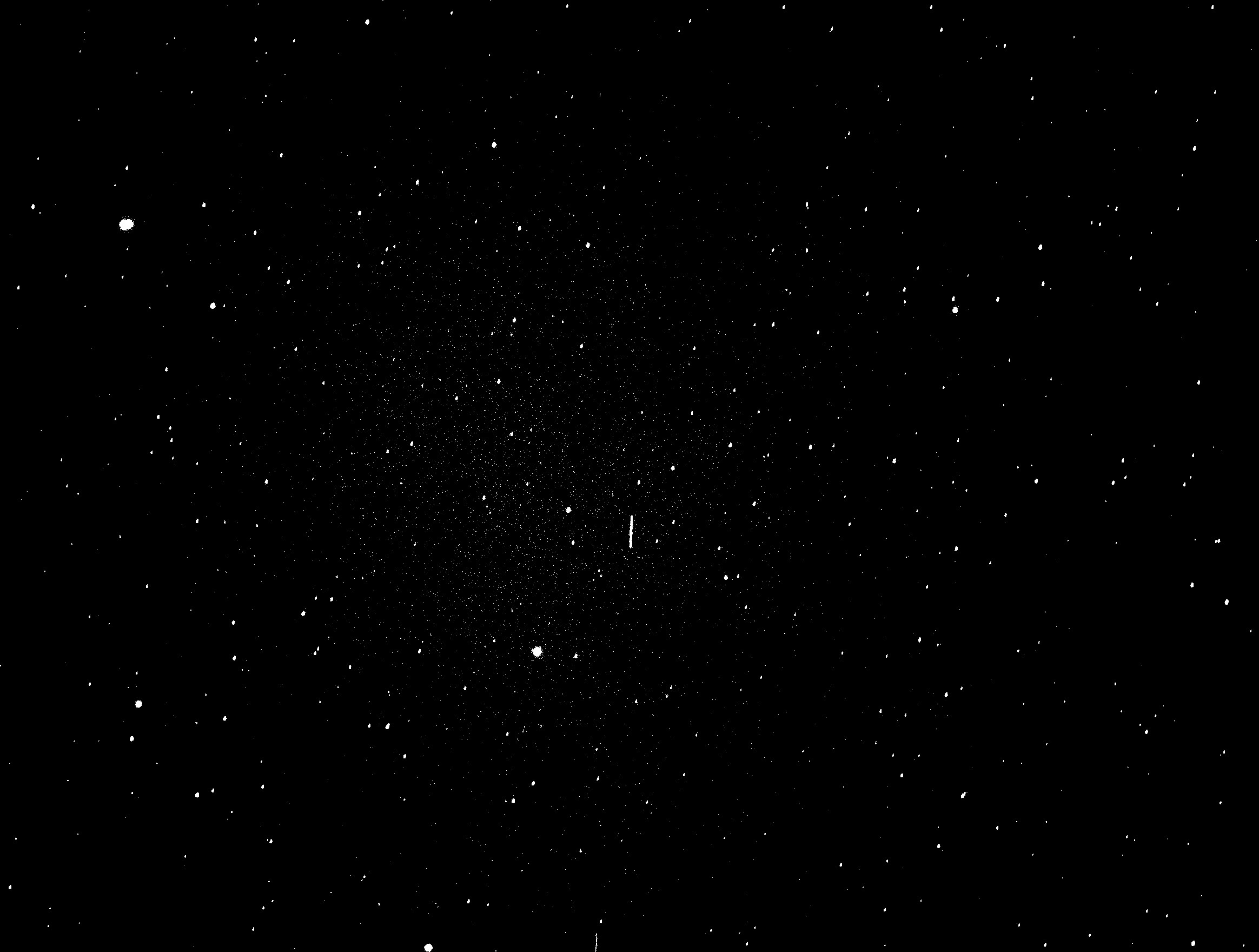}
    \end{minipage}
    \caption{Image processing example raw image is shown on the left and processed binary mapping is shown on the right.}
    \label{fig:binary:map}
\end{figure}

\begin{figure}[!htb]
    \centering
    \begin{minipage}{0.48\textwidth}
        \centering
        \includegraphics[width=0.9\textwidth]{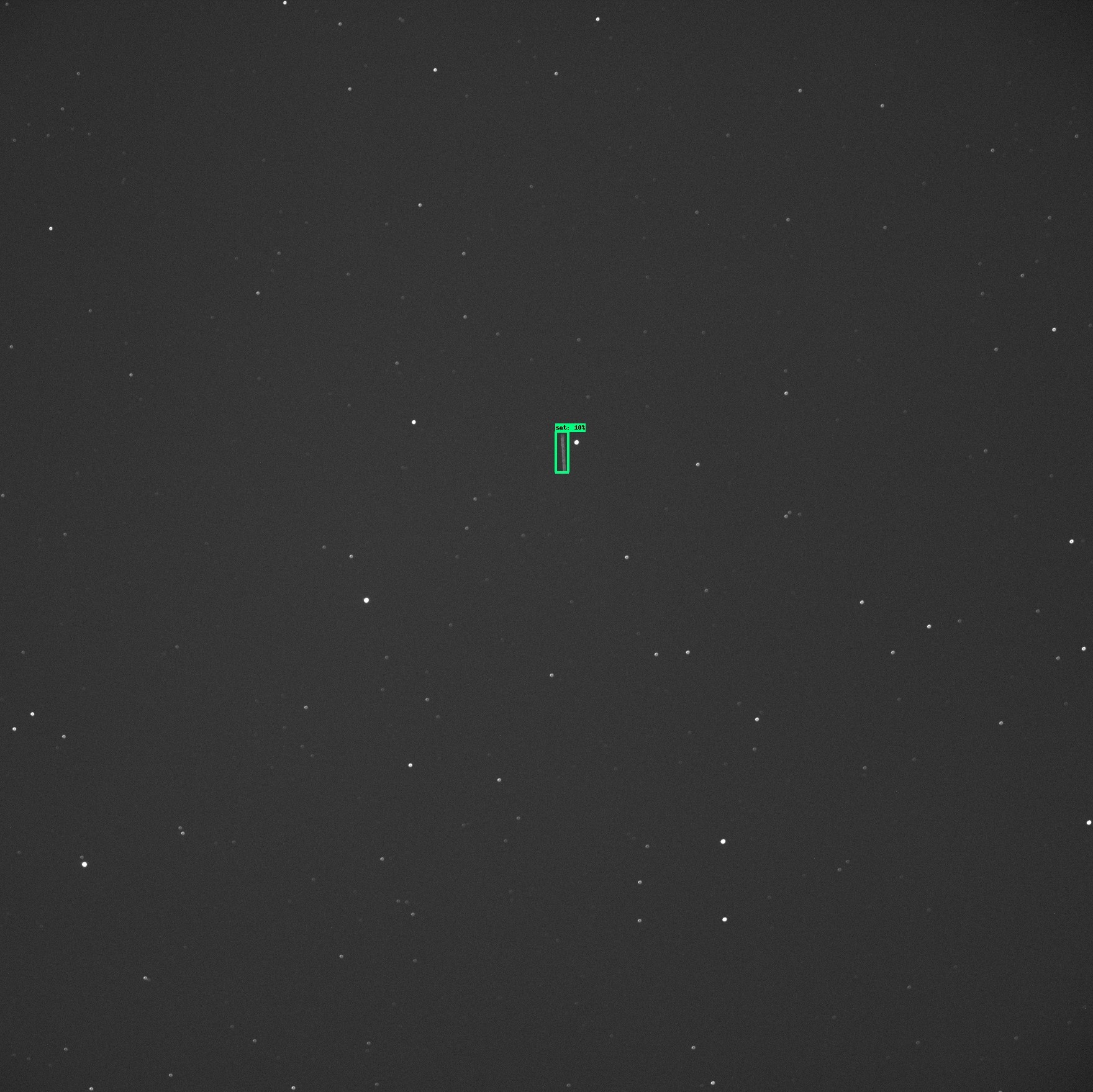}
    \end{minipage}%
    \begin{minipage}{0.48\textwidth}
        \centering
        \includegraphics[width=0.9\textwidth]{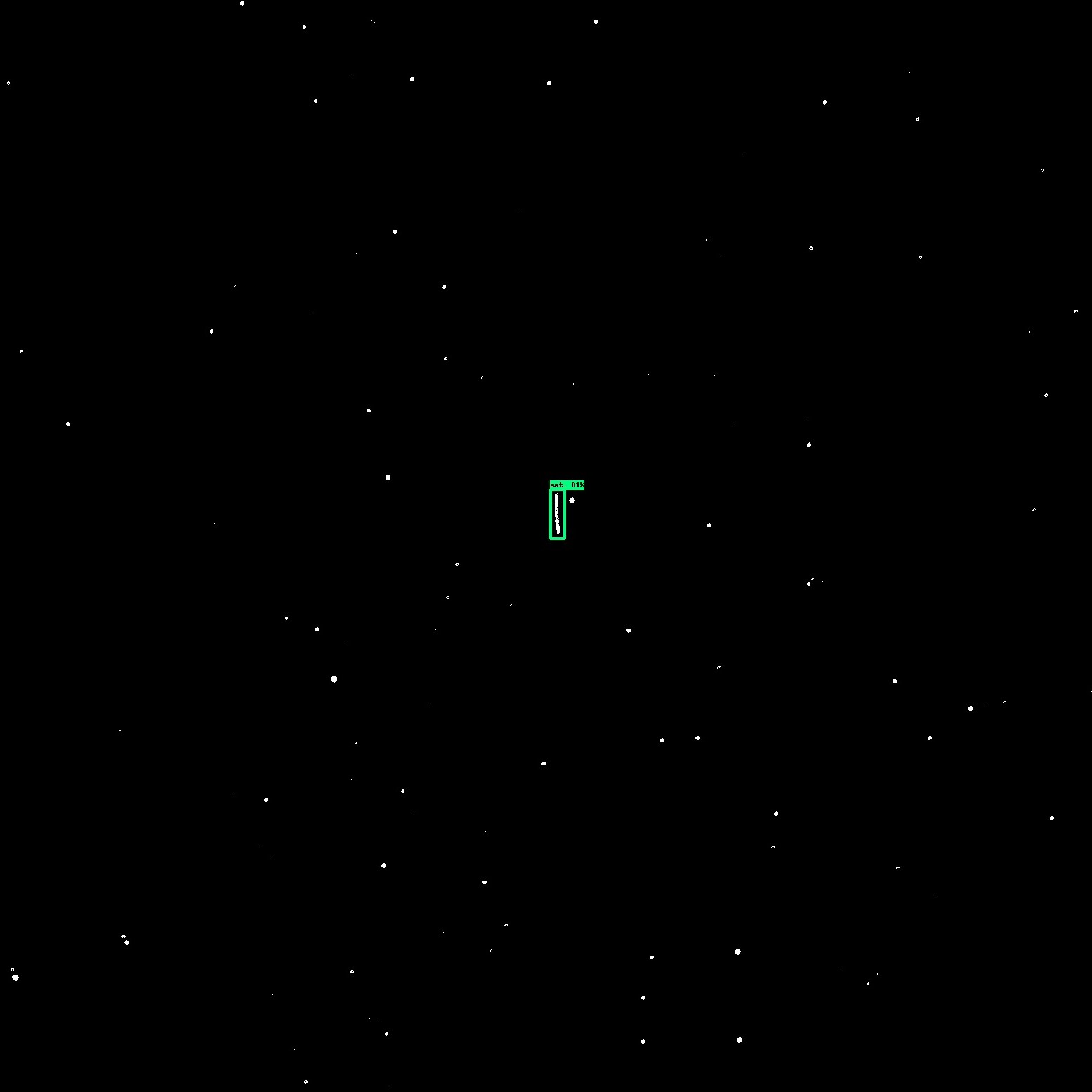}
    \end{minipage}
    \caption{Inference example on raw image (left) and processed image (right).}
    \label{fig:increased:inference}
\end{figure}

To increase model performance and inference confidence, raw images from the telescope camera were subjected to two stages of processing. The first was a conversion from a large raw file to a compressed jpg. Useful data from the image is mostly from brightness and darkness values, making any lost quality negligible. Consequently, because of the smaller size, image upload during inference is done more rapidly. The second stage converted the image to a binary black-and-white map. Any pixels above a certain brightness were set to a value of one, while those below were set to be zero (corresponding to black). This last processing step reduced the file size significantly, dropping most pixels down to zero. As an added effect, the CNN was more easily able to detect the RSO outline in the image. An example raw image is shown on the left in Fig. \ref{fig:binary:map}. The corresponding processed binary map image is shown on the right in Fig. \ref{fig:binary:map}. The CNN's ability to detect an RSO was improved by the binary processing and the results are shown in Figure \ref{fig:increased:inference}. On the left inference was run on a raw image and a confidence of $10\%$. With processing applied, the result is shown on the right of Fig. \ref{fig:increased:inference} and the confidence increased to $81\%$. 

\section{Results} \label{sec:Results}
\subsection{Overview of Detection Results}
\begin{figure}[!htb]
    \centering
        \centering
    \begin{minipage}{0.5\textwidth}
        \centering
        \includegraphics[width=0.95\textwidth]{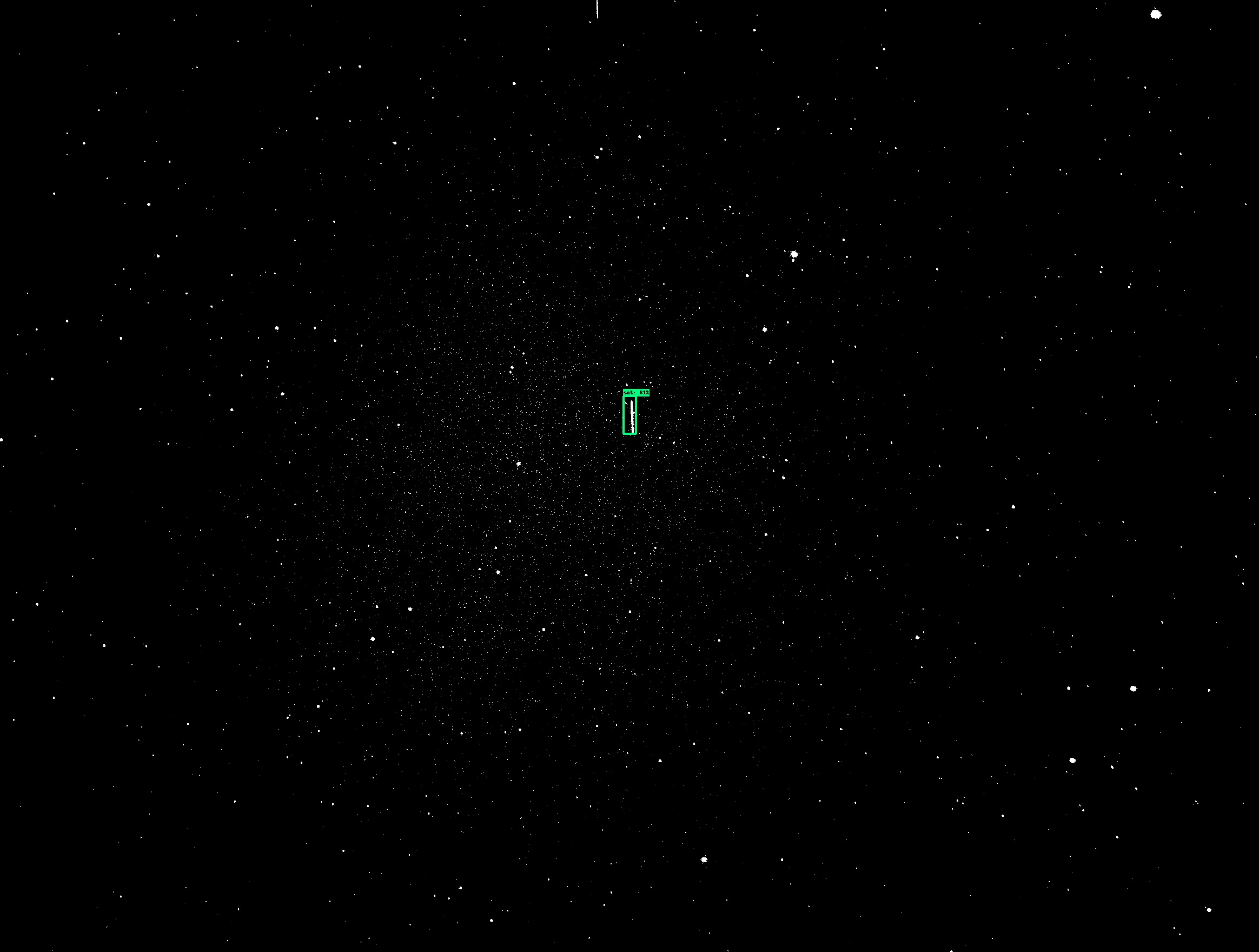}
    \end{minipage}%
    \begin{minipage}{0.5\textwidth}
        \centering
        \includegraphics[width=0.95\textwidth]{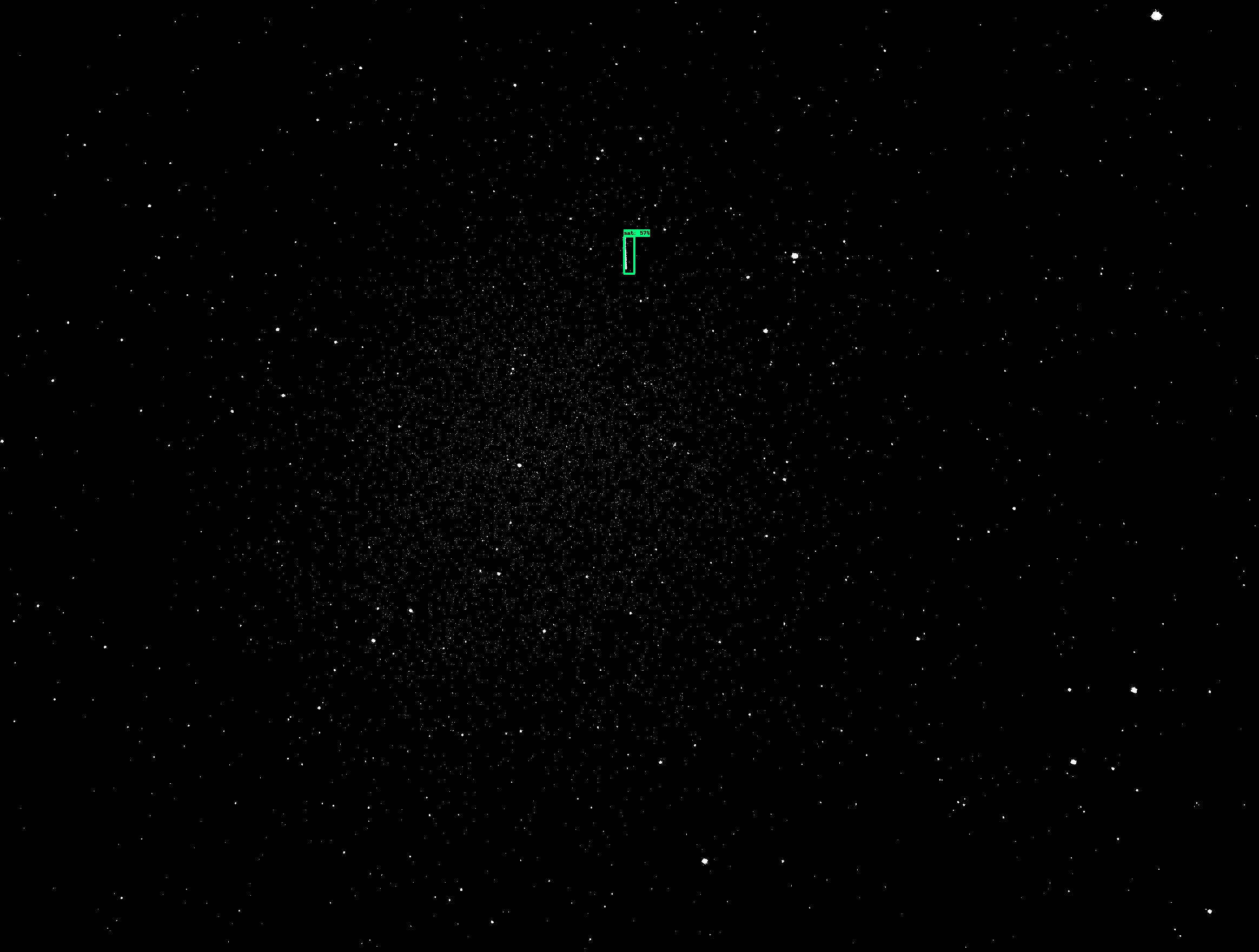}
    \end{minipage}
        \centering
    \begin{minipage}{0.5\textwidth}
        \centering
        \includegraphics[width=0.95\textwidth]{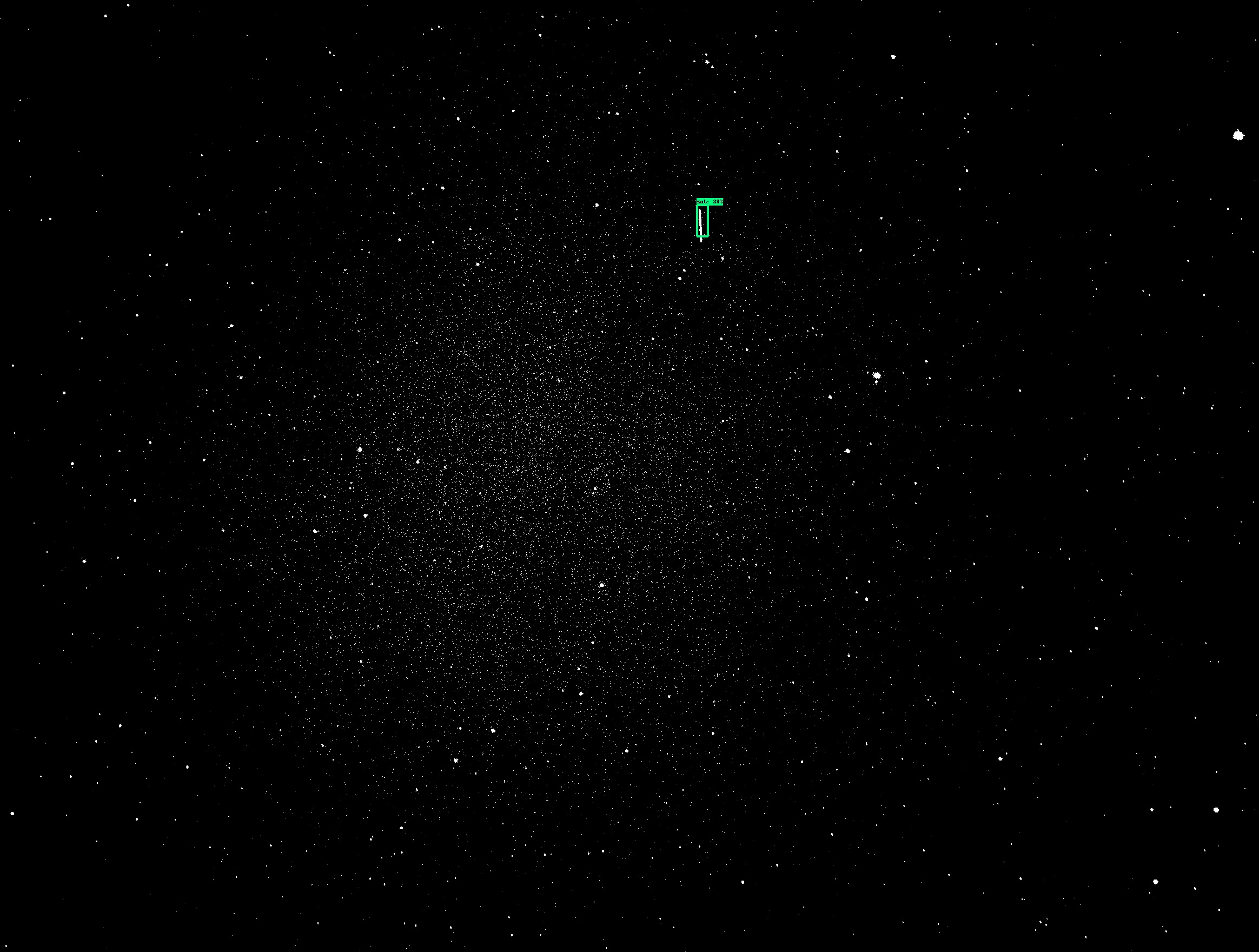}
    \end{minipage}%
    \begin{minipage}{0.5\textwidth}
        \centering
        \includegraphics[width=0.95\textwidth]{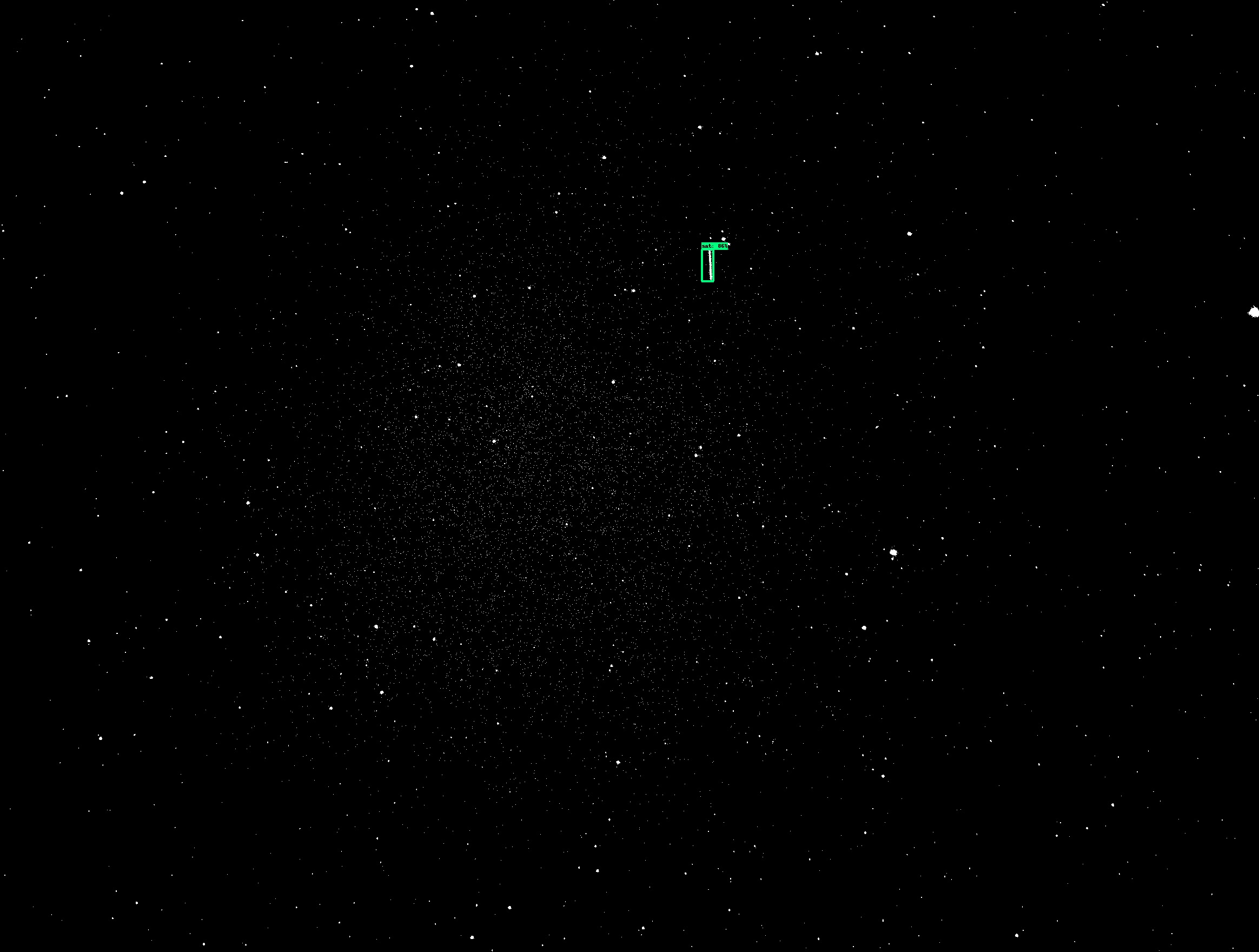}
    \end{minipage}
    \caption{Still capture sequence of the RSO detection program during operation. First frame is top left, continuing to top right, and final frame is shown in the bottom right.}
    \label{fig:observation:test}
\end{figure}

The performance of the system was gauged on its behavior and effectiveness in detecting and tracking satellites in Geostationary orbit (GEO). Test runs were conducted on multiple satellites to accommodate a variety of brightness levels and conditions. Figure \ref{fig:observation:test} shows imagery from one of these tests where the program successfully detected and continued tracking the satellite within the image frame. This can be seen as the RSO remained in view as the star field moved downward out of frame.

\begin{figure}[!htb]
    \centering
        \centering
    \begin{minipage}{0.5\textwidth}
        \centering
        \includegraphics[width=0.95\textwidth]{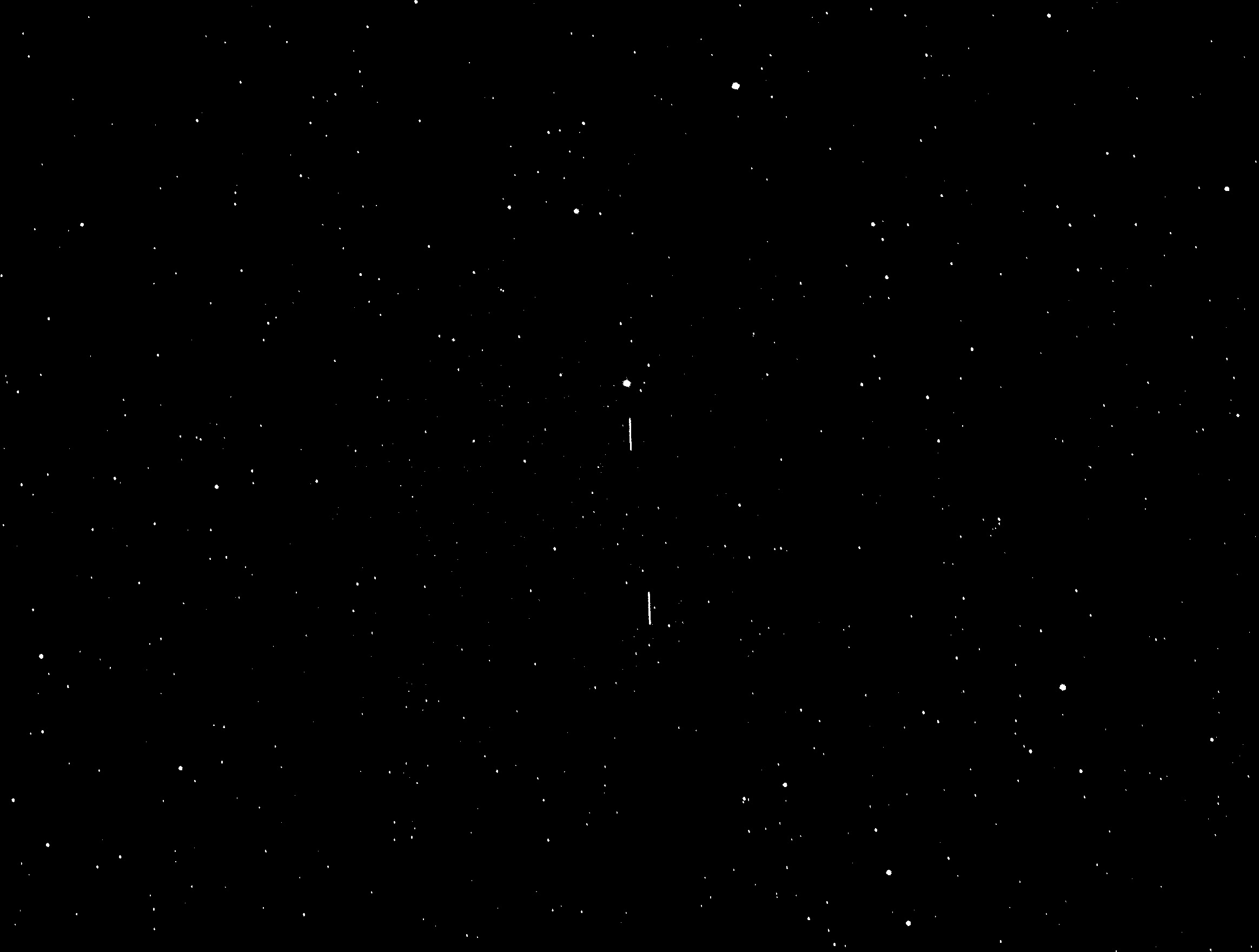}
    \end{minipage}%
    \begin{minipage}{0.5\textwidth}
        \centering
        \includegraphics[width=0.95\textwidth]{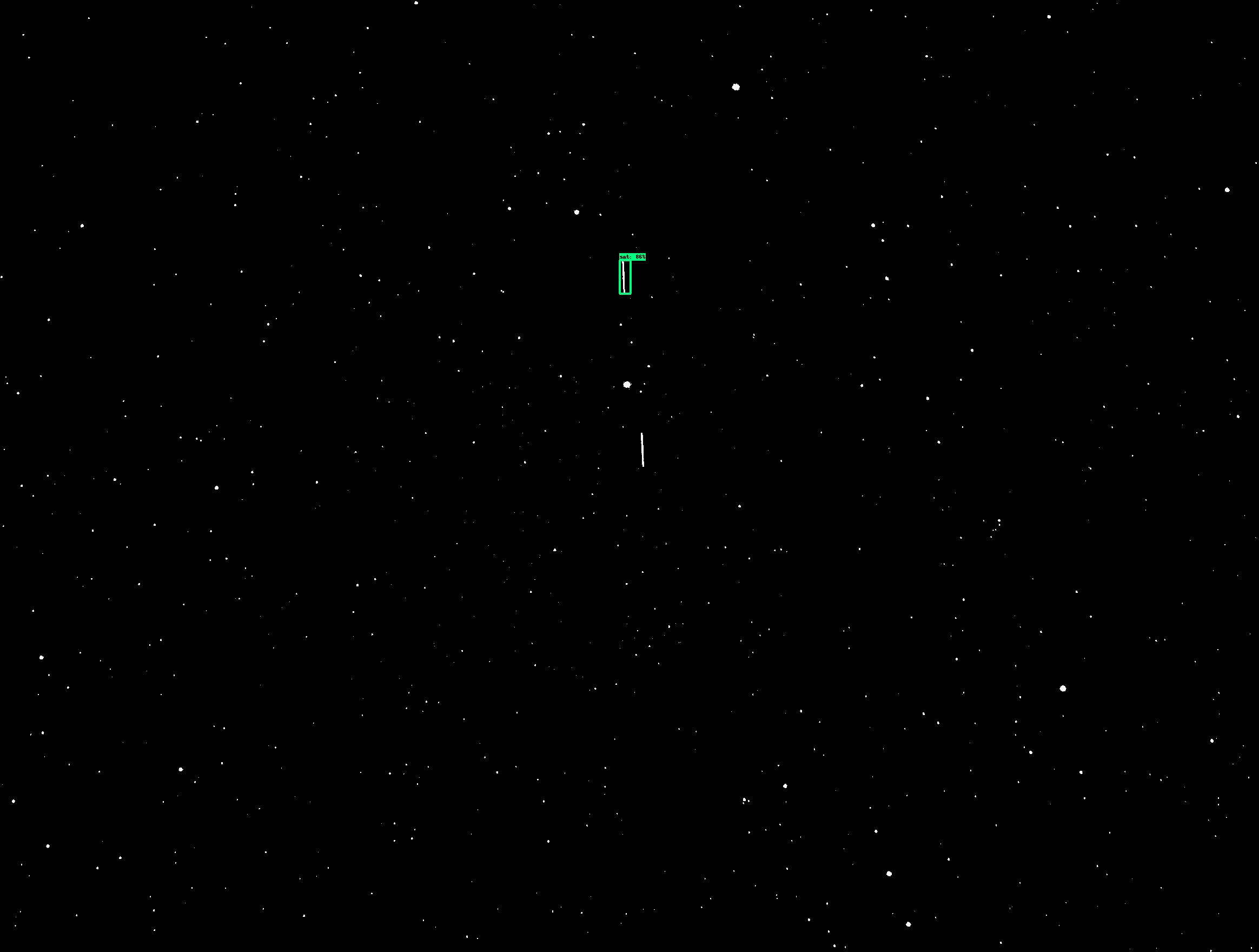}
    \end{minipage}
        \centering
    \begin{minipage}{0.5\textwidth}
        \centering
        \includegraphics[width=0.95\textwidth]{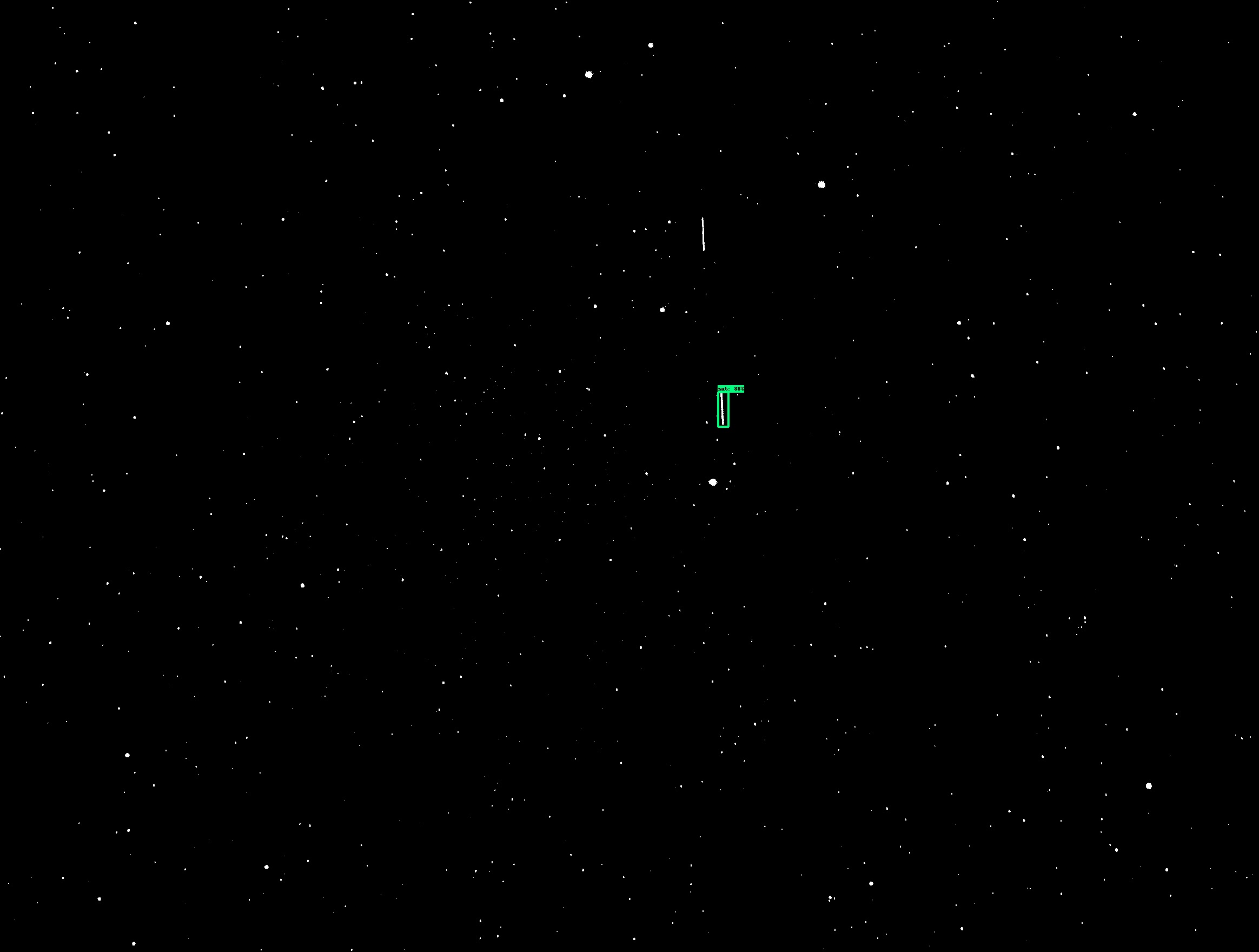}
    \end{minipage}%
    \begin{minipage}{0.5\textwidth}
        \centering
        \includegraphics[width=0.95\textwidth]{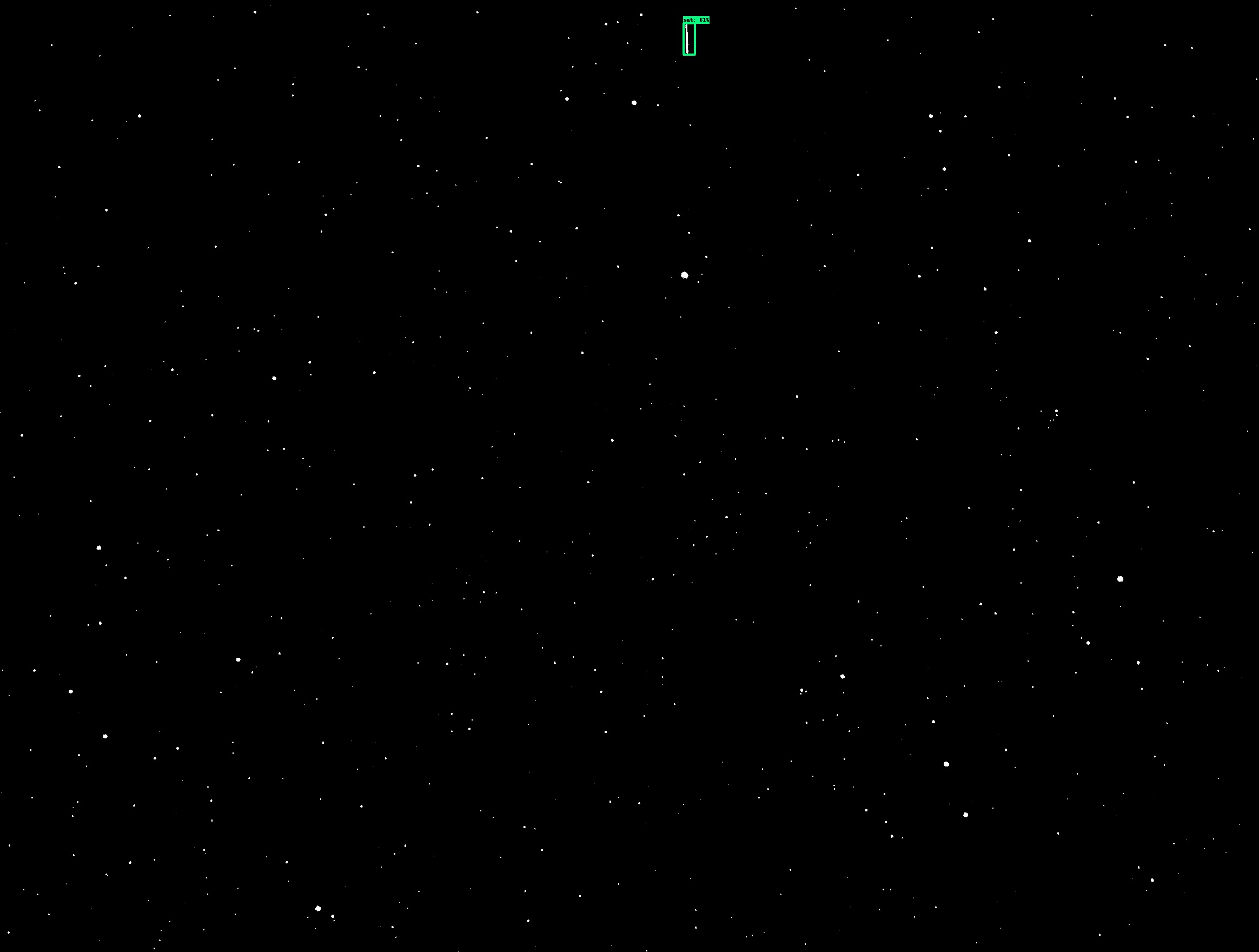}
    \end{minipage}
    \caption{Tracking of two RSOs in close proximity.}
    \label{fig:observation:double}
\end{figure}

The software also performs well when encountering clusters of RSOs within the image frame. This behavior can be seen in Figure \ref{fig:observation:double} where two satellites are tracked. 
While not specifically optimized for group detection, the software instead tracks the RSO with the highest detection in each frame and propagates. Since this is the first iteration of the full tracking program, it was manually limited to choosing only one detection to avoid possible duplication of centroid data and otherwise conflicting information which would complicate the experimental pipeline. Detection was affected by weather as we wanted to test the robustness of the system with different possible cases. Ideally, for a dedicated astronomy and observational night, atmospheric weather is chosen to be ideal (no clouds, low temperature), however, this might not be the case always. Some detections were affected by clouds as they lower the intensity of the RSO and stars requiring to change either integration time or just omitting the measurement. The network in this case with clouds was able to resolve the RSOs and identify them even with low confidence ($\approx10\%$). This still allowed the algorithm to continue processing and propagate correctly towards the RSO moving direction. Figure \ref{fig:confidence:intelsat-21} illustrates this phenomenon as a couple of clouds traversed through the sky in the camera's field of view.

\subsection{Autonomous Tracking Results}
This section shows the autonomous tracking performance of the LQE system used to predict an RSO's position and move the telescope in position to maintain visibility of the RSO in the telescope's field of view (FOV). Three different scenarios are presented tracking three different geostationary (GEO) satellites. Exposure times were constant and set at 10 seconds for all images. The threshold on detection confidence for the CNN was set at $10\%$ and produced no false positives during any of the tracking experiments. 

\begin{figure}[!htb]
    \centering
        \centering
    \begin{minipage}{0.49\textwidth}
        \centering
        \includegraphics[width=0.95\textwidth]{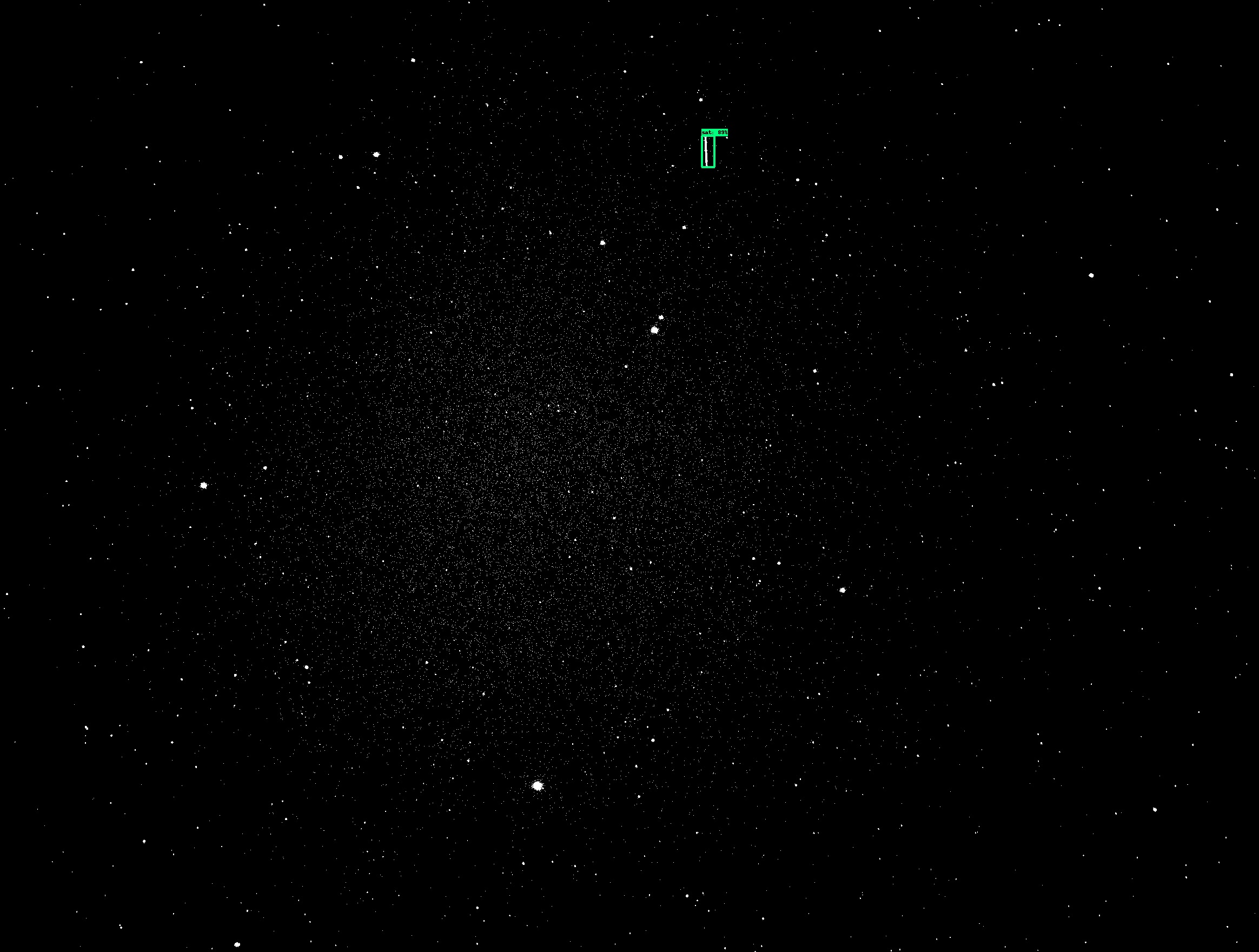}
        \caption{Example CNN detection result of Intelsat-14 with $89\%$ confidence. Detection is highlighted in a green box.}
        \label{fig:intelsat-14:output}
    \end{minipage}
    \hfill
    \begin{minipage}{0.49\textwidth}
    \centering
    \includegraphics[width=\textwidth]{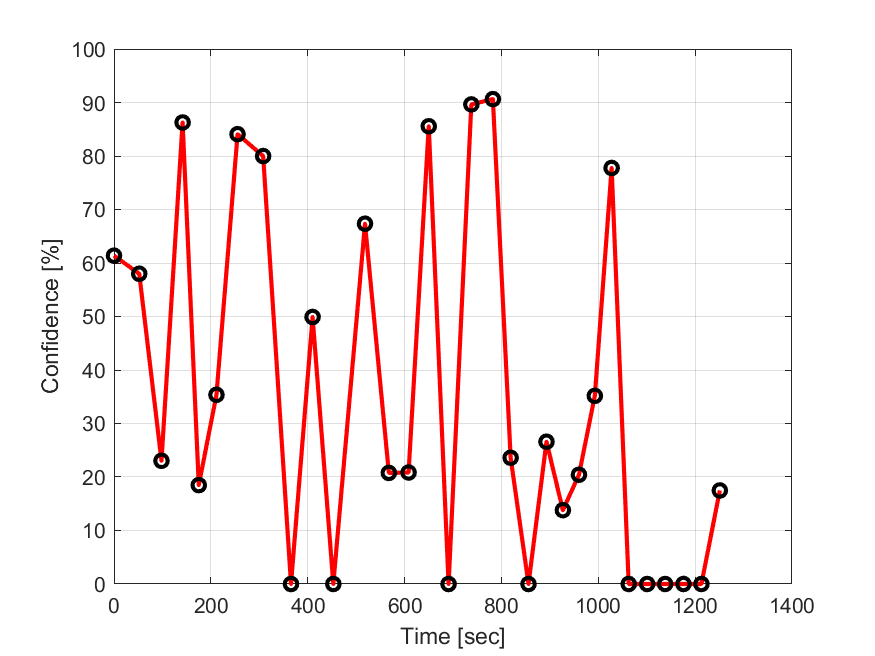}
    \caption{CNN detection confidence for Intelsat-14. Note that the satellite was not detected by the CNN intermittently through tracking and then was lost for several frames at the end of the dataset.}
    \label{fig:confidence:intelsat-14}
    \end{minipage}%
    \hfill
    \begin{minipage}{0.49\textwidth}
    \centering
    \includegraphics[width=\textwidth]{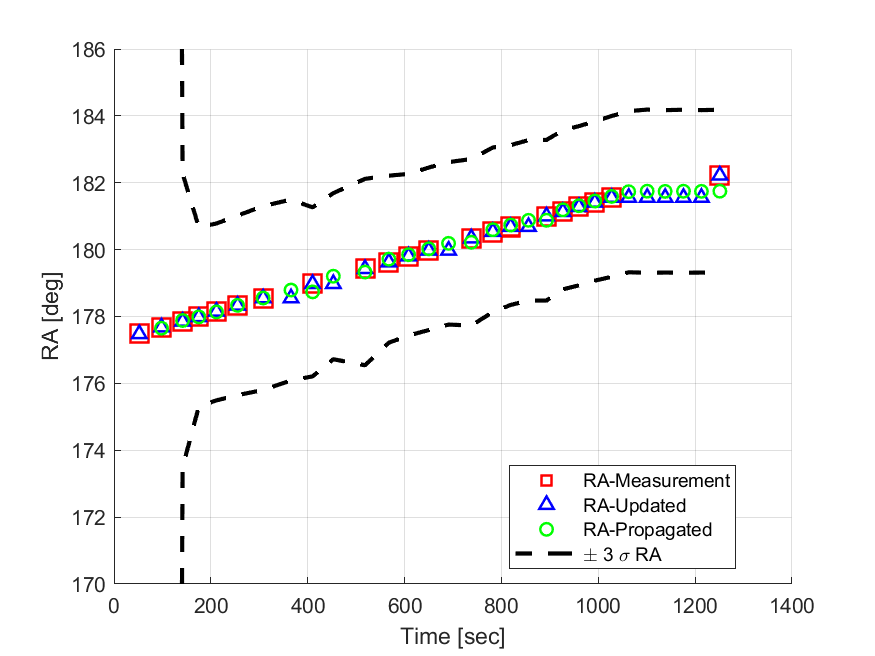}
    \caption{Tracking filter results in RA for Intelsat-14. Note that missing measurements are handled by the filter continuing to propagate from the last detected positions.}
    \label{fig:raplot:intelsat-14}
    \end{minipage}
    \begin{minipage}{0.5\textwidth}
    \centering
    \includegraphics[width=\textwidth]{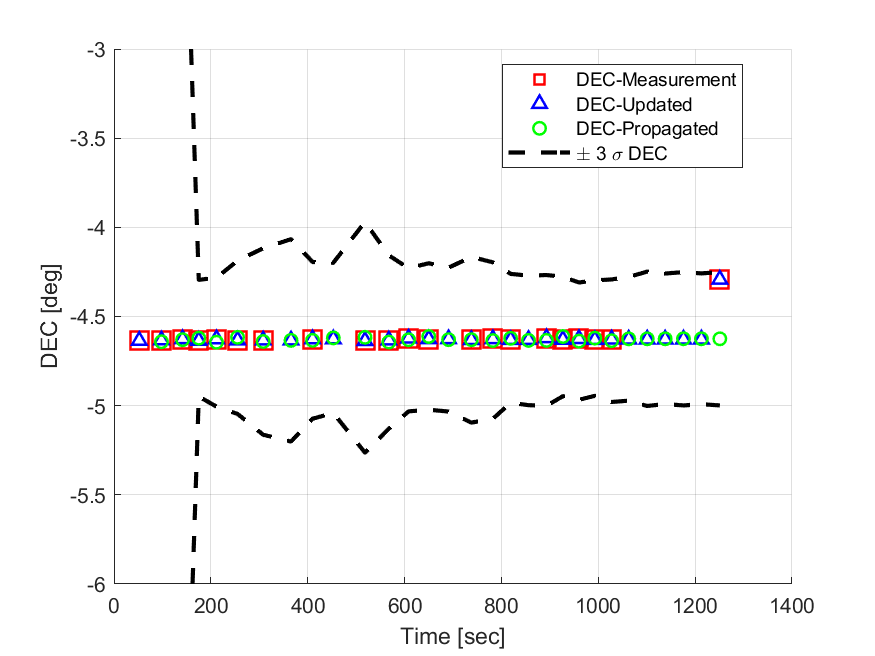}
    \caption{Tracking filter results in DEC for Intelsat-14.}
    \label{fig:decplot:intelsat-14}
    \end{minipage}
\end{figure}

The first scenario tracked the satellite Intelsat-14 (NORAD ID 36097)\cite{intelsat14-oh} a communications satellite that services North and South America as well as parts of Europe and Africa. In order to test the autonomous detection system's ability to initialize tracking on its own, the telescope was pointed such that the RSO would be in the field of view to kick off the tracking algorithm. Once two-successful detections occur in a row the LQE filter was initialized and used to predict the RSOs next location to send a slew command to the telescope. All images were taken with an exposure time of 10 seconds allowing for significant streaking of RSO objects while the telescope was set to track at a sidereal rate. An example image showing the CNN detection of Intelsat-14 is shown in Fig. \ref{fig:intelsat-14:output}. The confidence level of CNN detections across all images while tracking Intelsat-14 are shown in Fig. \ref{fig:confidence:intelsat-14}. A handful of missed detections can be seen in Fig. \ref{fig:confidence:intelsat-14}, however, the LQE filter was able to keep track of the RSO's inertial position, and the RSO was reacquired in later frames. Figures \ref{fig:raplot:intelsat-14} and \ref{fig:decplot:intelsat-14} show the filter and measured RA and DEC for Intelsat-14 for the duration tracking was performed. 

Note that the measurements from the RSOs detected position by the CNN are plotted as red boxes. The filter's position updated with the measurement is shown as a blue triangle, and the filter's predicted position for the next time step is shown as a green circle. The $3-\sigma$ bounds of the filter covariance for RA and DEC are shown as dashed black lines. Notably missed measurements did not cause the filter to lose track of the RSOs position, additionally, all measurements and predictions lie within the $3-\sigma$ bounds. The high value of the covariance dashed lines for the first 100 seconds results from the initialization values used for covariance on RA and DEC. Once the filter begins receiving measurements the covariance values for both RA and DEC converge to near-constant bounds around the true values. 

\begin{figure}[!htb]
    \centering
        \centering
    \begin{minipage}{0.49\textwidth}
    \centering
    \includegraphics[width=0.95\textwidth]{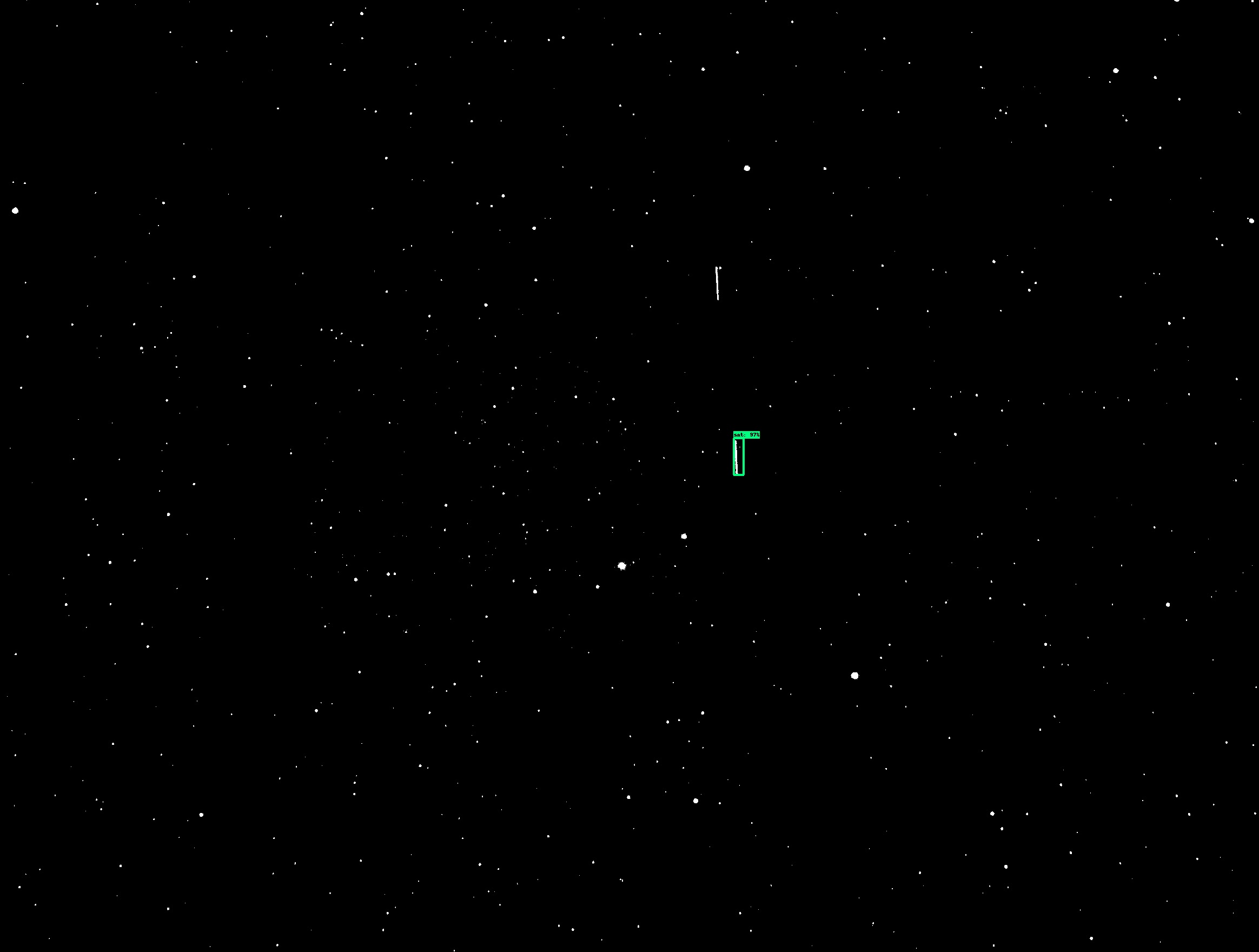}
    \caption{Example CNN detection result of SES-10 with $97\%$ confidence. Detection is highlighted in a green box.}
    \label{fig:ses-10:output}
    \end{minipage}
    \hfill
    \begin{minipage}{0.49\textwidth}
    \centering
    \includegraphics[width=\textwidth]{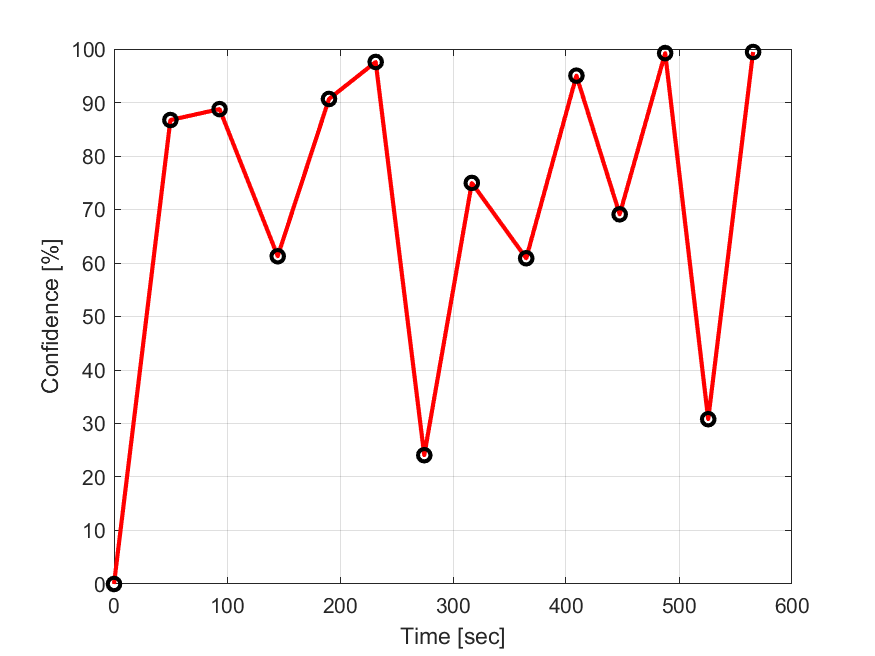}
    \caption{CNN detection confidence for SES-10.}
    \label{fig:confidence:ses-10}
    \end{minipage}
    \begin{minipage}{0.49\textwidth}
    \centering
    \includegraphics[width=\textwidth]{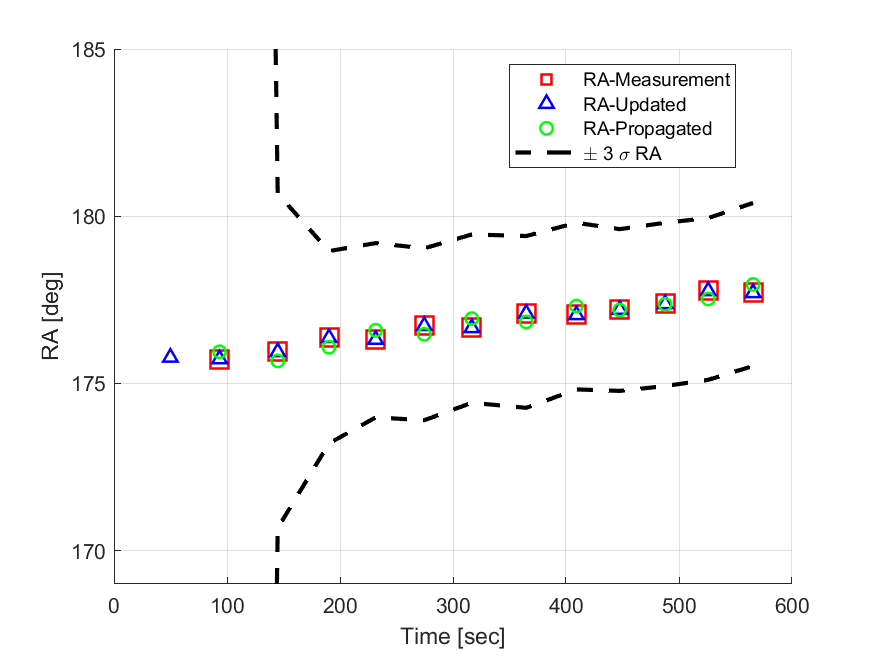}
    \caption{Tracking filter results in RA for SES-10.}
    \label{fig:raplot:ses-10}
    \end{minipage}
    \begin{minipage}{0.49\textwidth}
    \centering
    \includegraphics[width=\textwidth]{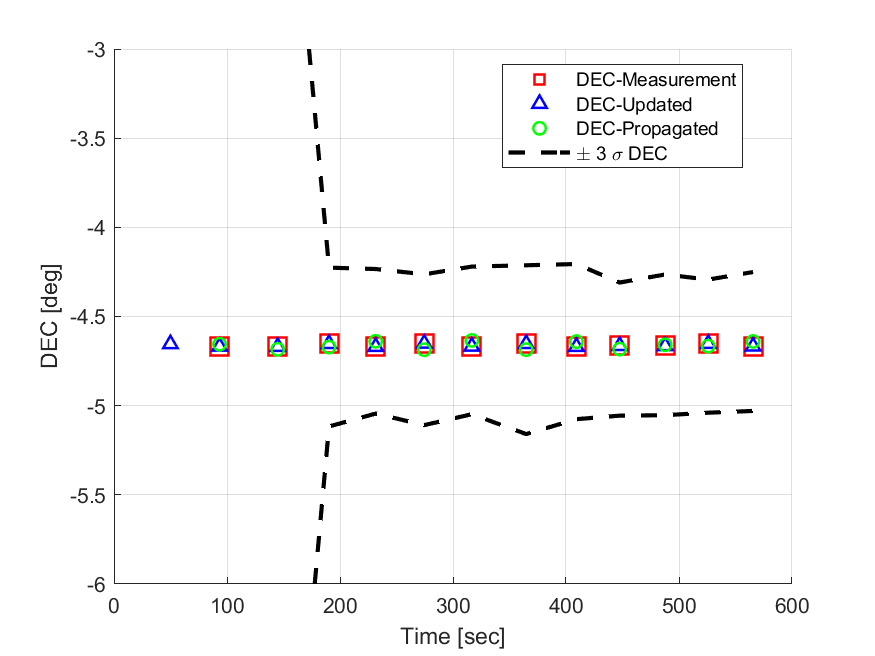}
    \caption{Tracking filter results in DEC for SES-10.}
    \label{fig:decplot:ses-10}
    \end{minipage}
\end{figure}

The second scenario used to test the autonomous tracking system was tracking of the GEO satellite SES-10 (NORAD ID: 42432), a Ku band communications satellite.\cite{SES-10-wr} The same approach as the first scenario was used to initialize tracking. The telescope was manually moved into a position where the satellite SES-10 would be in the telescope's FOV so that the CNN would begin detections and kick off the LQE tracker. Once two successful detections were made, the first two measurements were once again used to initialize the LQE tracker. An example of the CNN detection of SES-10 is shown in Fig. \ref{fig:ses-10:output}. Note that two satellite streaks are visible as a neighboring GEO satellite is in the same telescopic FOV. This test scenario corresponds to the example shown above in Fig. \ref{fig:observation:double} of tracking two RSOs in close proximity. The highest confidence resulting detection was used as the filter update measurement. The confidence of detection for each image frame is shown in Fig. \ref{fig:confidence:ses-10}. The filter results for RA and DEC are shown in Figs. \ref{fig:raplot:ses-10} and \ref{fig:decplot:ses-10} respectively. The same filter behavior is observed where the covariance takes some time (about 200 seconds) to converge to a stable value. No measurements or predicted values lie outside the $3-\sigma$ bounds of the filter covariance. 

\begin{figure}[!htb]
    \centering
        \centering
    \begin{minipage}{0.49\textwidth}
    \centering
    \includegraphics[width=0.95\textwidth]{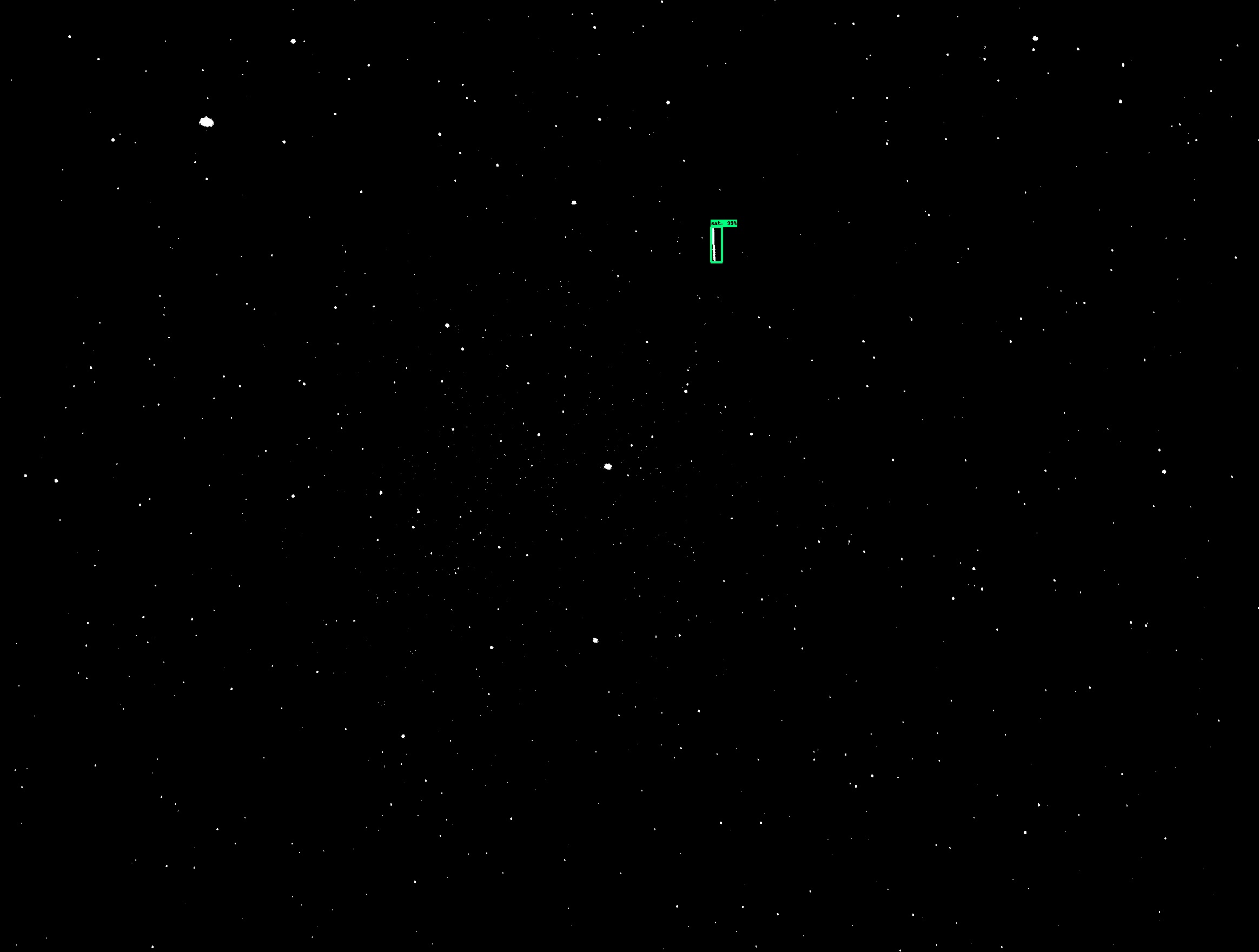}
    \caption{Example CNN detection result of Intelsat-21 with $99\%$ confidence. Detection is highlighted in a green box.}
    \label{fig:intelsat-21:output}
    \end{minipage}
    \hfill
    \begin{minipage}{0.49\textwidth}
    \centering
    \includegraphics[width=\textwidth]{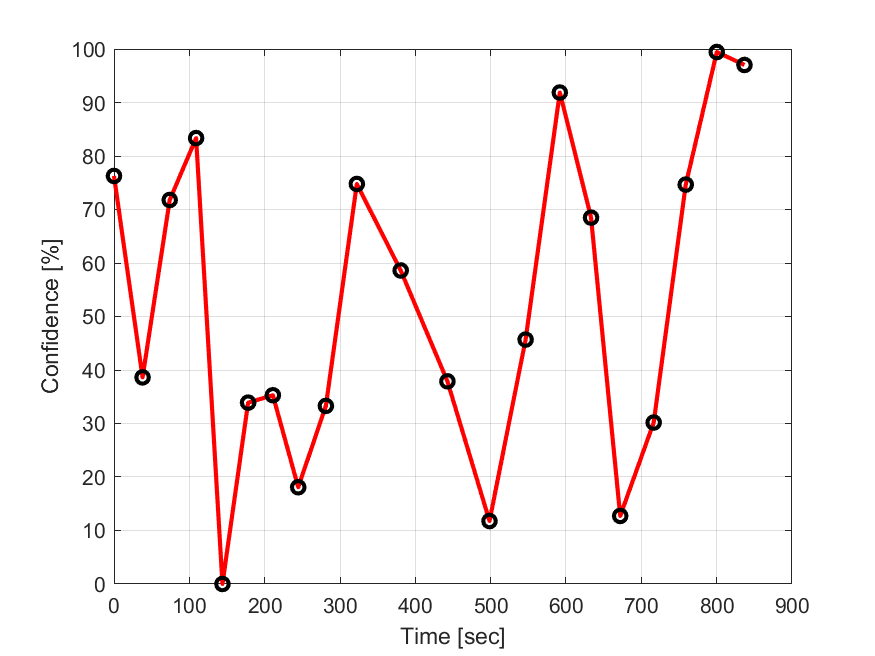}
    \caption{CNN detection confidence for Intelsat-21.}
    \label{fig:confidence:intelsat-21}
    \end{minipage}
    \begin{minipage}{0.49\textwidth}
    \centering
    \includegraphics[width=\textwidth]{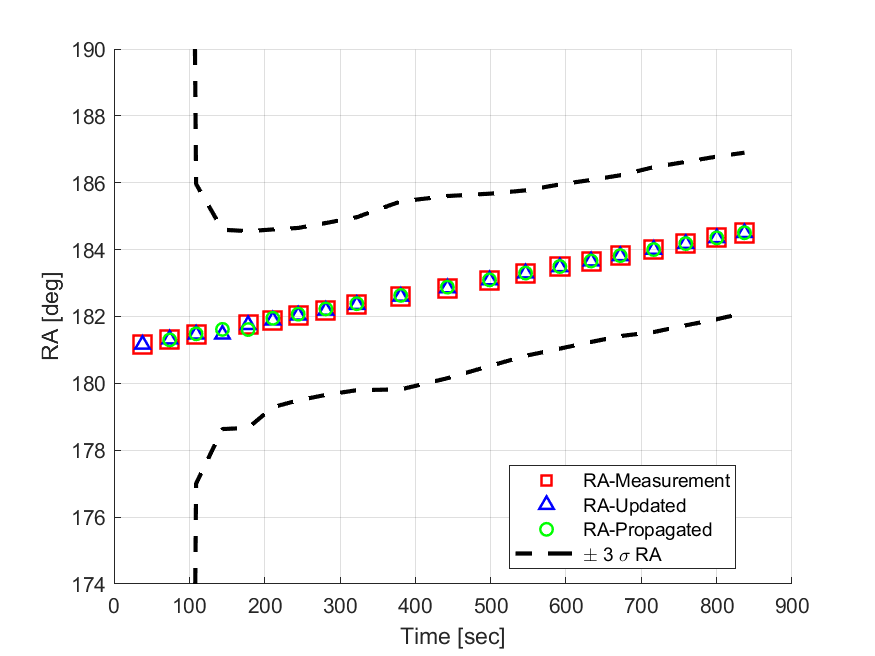}
    \caption{Tracking filter results in RA for Intelsat-21.}
    \label{fig:raplot:intelesat-21}
    \end{minipage}
    \begin{minipage}{0.49\textwidth}
    \centering
    \includegraphics[width=\textwidth]{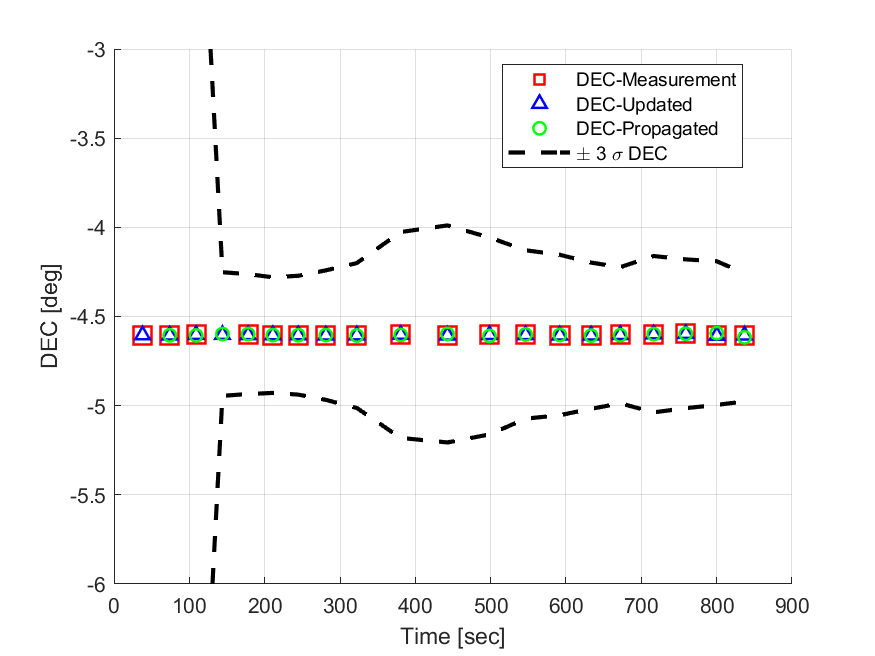}
    \caption{Tracking filter results in DEC for Intelsat-21.}
    \label{fig:decplot:intelsat-21}
    \end{minipage}
\end{figure}

The final scenario used for testing the autonomous detection framework was a tracking scenario for Intelsat-21 (NORAD ID: 38749) a C-band communications GEO satellite.\cite{intelsat21-fz} An example image showing the CNN detection of Intelsat-21 is given in Fig. \ref{fig:intelsat-21:output}. CNN detection confidence for each image in the test sereies is shown in Fig. \ref{fig:confidence:intelsat-21}. The RA and DEC filter results are shown in Figures \ref{fig:raplot:intelesat-21} and \ref{fig:decplot:intelsat-21} respectively. There was one missed detection by the CNN at around 150 seconds into the scenario, however, the missed detection did not affect the overall tracking performance. As with the previous two scenarios, the filter covariance values converged in about 150 seconds to stable values.

\section{Orbit Estimation}
The inertial observations of an RSO tracked by the system can be utilized for orbit estimation. One way to solve such a problem can be by using the application of the modified Gooding algorithm for initial orbit determination (IOD) as described by Henderson et al.\cite{henderson2010modifications} This modification allows to improve the IOD with more measurements and then it can be propagated using any type of propagator such as Runge-Kutta 4. A future version of the autonomous tracking algorithm will include an orbit estimation component to develop an online estimate of the object being tracked.

\section{Conclusions and Future Work}
The system proved effective at tracking several GEO satellites without prior knowledge of their motion. The CNN architecture was successful at performing inference on processed images captured in real-time utilizing a low-power/cost Raspberry Pi computer. Improvements to the system should focus on decreasing the run-time of several operations including, image processing, and inference. Cutting down on processing time will decrease the time between measurements and enable better tracking with the LQE system. This can be further augmented by using an embedded device with GPU acceleration or various parallelization techniques which streamline the pipeline. Future work also includes testing on a wider variety of satellite orbits and the ability to track RSO clusters as a group. With optimized run-time, the system can be adapted to track orbits that would require a faster update such as low-Earth-orbit (LEO) objects. Additionally, an online orbit estimation method should be incorporated to estimate the observed RSOs orbit in real-time.

\acknowledgments 
This work was partially supported by the National Defense and Science and Engineering Graduate (NDSEG) fellowship program. 

\bibliography{QuasiRealTime} 
\bibliographystyle{spiebib} 

\end{document}